\documentclass[lettersize,journal]{IEEEtran}

\usepackage[inline]{enumitem}
\usepackage{pifont}

\usepackage{graphicx}
\usepackage{textcomp}
\usepackage{xcolor}
\usepackage{CJKutf8}

\usepackage{stfloats} % 必须加！修复双栏浮动体顺序bug
\usepackage{amsmath,bm}
\usepackage{amssymb}
\usepackage{amsthm}
\usepackage{amsfonts}

\usepackage{bm}
\usepackage{relsize}
\usepackage{cases}
\usepackage{caption}

\usepackage{multirow}
\usepackage{tabularx}
\usepackage{hhline}
\usepackage{diagbox}

\usepackage[cal=boondoxupr]{mathalpha}
\usepackage[para,online,flushleft]{threeparttable}
\usepackage{listings} %导入包
\usepackage{soul}  
\usepackage{xcolor}
\usepackage{booktabs}
\usepackage{setspace} % 调整行间距

\usepackage{overpic, pict2e}
\usepackage{pgfplots}
\pgfplotsset{compat=1.16}
\usepackage[caption=false,font=footnotesize]{subfig}
\usepackage{algorithm}

\IfFileExists{algpseudocodex.sty}{%
\RequirePackage[italicComments=false, %commentColor=blue, 
    rightComments=false, endLComment=, beginComment=//~]{algpseudocodex}%
}{%
\PackageWarning{algpseudocode}{%
    Package algpseudocodex not available, therefore\MessageBreak
    substituting command LComment for Comment\MessageBreak
}%
\newcommand*{\LComment}{\Comment}%
}%

\usepackage{cite}
\usepackage[hyphens]{url}
\usepackage[normalem]{ulem}
\usepackage[bookmarks=true,breaklinks=true,letterpaper=true,colorlinks,citecolor=black,linkcolor=black,urlcolor=black]{hyperref}

\newcommand{\todo}[1]{}

\newcommand*{\cref}[1]{\S\ref{#1}}
\newcommand*{\Cref}[1]{\S\ref{#1}}

\makeatletter
\def\IEEElabelanchoreqn#1{\bgroup
\def\@currentlabel{\p@equation\theequation}\relax
\def\@currentHref{\@IEEEtheHrefequation}\label{#1}\relax
\Hy@raisedlink{\hyper@anchorstart{\@currentHref}}\relax
\Hy@raisedlink{\hyper@anchorend}\egroup}
\makeatother

\definecolor{outlinetextcolor}{RGB}{171,137,240}

\usepackage{tcolorbox}

\usepackage{cases}
\usepackage{tikz}
\newcommand{\mytikzmark}[1]{\tikz[overlay,remember picture] \node[inner sep=0pt, outer sep=0pt] (#1) {};}

\usepackage{pbox}

\newif\ifrevision
\revisionfalse

\NewDocumentCommand{\revision}{O{blue} o m}{%
  \ifrevision
    {\color{#1}#3}%
    \IfValueT{#2}{\marginpar{\scriptsize\sffamily\color{#1}[to #2]}}%
  \else
    #3%
  \fi
}
\NewDocumentCommand{\revisioninline}{O{blue} o m}{%
  \ifrevision
    \IfValueT{#2}{{\footnotesize\sffamily\color{#1}[#2]~}}%
    {\color{#1}#3}%
  \else
    #3%
  \fi
}

\usepackage{xspace}

\newcommand\ETEs{E2E\xspace}
\newcommand\ETEB{E2E\xspace}
\newcommand*{\HWName}{Tile-based\xspace}
\newcommand*{\hwname}{tile-based\xspace}
\newcommand*{\aux}{cockpit\xspace}

\newcommand*{\singlefunc}{task\xspace}
\newcommand*{\singlefuncs}{tasks\xspace}

\newcommand*{\ADSArch}{DNN-based\xspace}

\newcommand{\execTime}{execution time\xspace}

\newcommand{\PaperName}{{ADS-Tile}\xspace}

\newcommand*{\CyclicName}{Cyc.\xspace}
\newcommand*{\TpFirstName}{Tp-driven\xspace}

\DeclareBoldMathCommand{\set}{set}

\newcommand*{\slack}{\mathcal{D}} 

\renewcommand*{\xi}{\mathit{chain}}

\newcommand*{\Tile}{tile}

\usepackage{xparse}

\newcommand*{\PartitionMark}{partition} %\mathbb{P}
\newcommand*{\PartitionNum}{N_{\PartitionMark}}

\NewDocumentCommand{\cfg}{o m O{}}{%
    \IfValueTF{#1}
    {imp_{#2,#1}^{#3}}
    {imp_{#2}^{#3}}
}

\newcommand*{\Xcross}{\ding{55}}

\newcommand{\AbbrCallout}[1]{\textsf{#1}}
\newcommand{\concept}[1]{\textbf{#1}} 
\newcommand{\option}[1]{\textsf{#1}} 
\newcommand{\highlight}[1]{\textsl{#1}} 
\newcommand{\highlightuline}[1]{\highlight{\uline{#1}}} 
\newcommand{\highlightit}[1]{\textit{#1}}

\usepackage{xcolor}
\definecolor{codebrown}{rgb}{0.8,0.44,0.2}
\definecolor{codegray}{rgb}{0.5,0.5,0.5}
\definecolor{codepurple}{rgb}{0.58,0,0.82}
\definecolor{backcolour}{rgb}{0.95,0.95,0.92}

\newcommand{\squishlist}{
 \begin{list}{$\bullet$}
  { \setlength{\itemsep}{0pt}
     \setlength{\parsep}{3pt}
     \setlength{\topsep}{3pt}
     \setlength{\partopsep}{0pt}
     \setlength{\leftmargin}{1.5em}
     \setlength{\labelwidth}{1em}
     \setlength{\labelsep}{0.5em} } }
\newcommand{\squishend}{
  \end{list}  }

\newenvironment{compacteqn}{%
  \par\begingroup\small
  \setlength{\abovedisplayskip}{3pt plus 1pt minus 1pt}%
  \setlength{\belowdisplayskip}{3pt plus 1pt minus 1pt}%
  \setlength{\abovedisplayshortskip}{2pt plus 1pt minus 1pt}%
  \setlength{\belowdisplayshortskip}{2pt plus 1pt minus 1pt}%
  \setlength{\jot}{1pt}%
}{%
  \par\endgroup
}

  \newcolumntype{M}{>{\centering\arraybackslash}m{2cm}} % Adjust the width as needed

\newif\ifcompactformat
\compactformattrue
\ifcompactformat
\renewcommand{\baselinestretch}{0.95} % 压缩行间距 (默认 1.0)

\setlist[itemize]{leftmargin=*, noitemsep, topsep=1pt, parsep=0pt, partopsep=0pt}
\setlist[enumerate]{leftmargin=*, noitemsep, topsep=1pt, parsep=0pt, partopsep=0pt}

\fi

\begin{document}

\title{
    Isolation-aware Scheduling Framework for \ADSArch End-to-End Autonomous Driving System on \HWName Accelerators
} 
% \author{Chenguang Zhang, 

% % \thanks{This paper was produced by the IEEE Publication Technology Group. They are in Piscataway, NJ.}% <-this % stops a space
% % \thanks{Manuscript submitted to IEEE TC on 2025-07-06}
% }

% Department of \[具体院系\], 
\author{
\IEEEauthorblockN{Chenguang Zhang\IEEEauthorrefmark{1}\IEEEauthorrefmark{2}, Yuanpeng Zhang\IEEEauthorrefmark{2}, Chenhao Xue\IEEEauthorrefmark{2}, Yihan Yin\IEEEauthorrefmark{2}, 
% Bo Yu\IEEEauthorrefmark{4}, Shaoshan Liu\IEEEauthorrefmark{4}, Yuan Xie\IEEEauthorrefmark{5}, 
Chen Zhang\IEEEauthorrefmark{6}, Guangyu Sun\IEEEauthorrefmark{2}\IEEEauthorrefmark{3}\IEEEauthorrefmark{7}}\thanks{\IEEEauthorrefmark{7}Corresponding author.}\\
% Funding statement can be enabled after the exact grants are confirmed:
\thanks{This work is supported by Beijing Natural Science Foundation (L243001), National Natural Science Foundation of China (U25A6023), and 111 Project (B18001).}
\IEEEauthorblockA{
\IEEEauthorrefmark{1}School of Computer Science, Peking University, Beijing, China\\
\IEEEauthorrefmark{2}School of Integrated Circuits, Peking University, Beijing, China\\
\IEEEauthorrefmark{3}Beijing Advanced Innovation Center for Integrated Circuits, Beijing, China\\
% \IEEEauthorrefmark{4}Shenzhen Institute of Artificial Intelligence and Robotics for Society, Shenzhen, China\\
% \IEEEauthorrefmark{5}HKUST, Hong Kong, China, 
\IEEEauthorrefmark{6}Shanghai Jiao Tong University, Shanghai, China\\
zhangchg@stu.pku.edu.cn, \{zyp\_cs, xch927027\}@pku.edu.cn, yyhsess2021@stu.pku.edu.cn, 
% \{boyu, shaoshanliu\}@cuhk.edu.cn, yuanxie@ust.hk, 
chenzhang.sjtu@sjtu.edu.cn, gsun@pku.edu.cn
}
}

\markboth{IEEE TRANSACTIONS ON COMPUTERS, VOL.~XX, NO.~XX, Month~Year}%
{Zhang \MakeLowercase{\textit{et al.}}: Isolation-aware Scheduling Framework for DNN-based End-to-End Autonomous Driving System on Tile-based Accelerators}


\maketitle
\begin{abstract}
    Level-4+ autonomous driving systems (ADS) must run dozens of heterogeneous
    deep neural networks (DNNs) as end-to-end (E2E) pipelines under a strict
    latency constraint (${\leq}100$\,ms), even as execution time varies by up to
    $3.3\times$. Cost rules out dedicating isolated hardware to each function in
    mass-produced ADS, so these DNNs must be densely colocated on a single chip,
    which introduces shared-resource contention. Tile-based accelerators expose
    two scheduling opportunities that conventional ADS schedulers do not exploit. First, they
    provide a tunable degree of parallelism (DoP): assigning more tiles raises DoP
    and can shorten DNN execution time. Second, they provide hardware-native
    isolation: tiles can be physically partitioned among co-located DNNs. But
    using this flexibility is expensive: changing a task's DoP triggers a
    stop-migrate-restart reallocation of its weights and intermediate features. At
    ADS task rates of 10--240\,Hz, these stalls accumulate along E2E chains and
    threaten deadlines. Reservation-based schedulers fix DoP and leave this
    flexibility unused; work-conserving schedulers exploit it but assume
    reallocation is cheap and treat deadlines as independent. We present ADS-Tile,
    an isolation-aware scheduling framework that targets the reallocation cost of
    spatial DoP changes. ADS-Tile combines configurable isolation and elastic
    reservation into a spatio-temporal isolation-sharing space that bounds where
    and when reallocation occurs; a probabilistic latency model and a DAG-aware
    runtime scheduler then use this space to decide task colocation and DoP under
    shared E2E deadlines. On an industry- and academia-derived ADS benchmark, ADS-Tile
    uses up to 32\% fewer tiles than the work-conserving baseline in
    deadline-critical settings and cuts reallocation-induced wasted processing
    capacity from 17\%--44\% to below 1.2\%. Controlled spatio-temporal sharing
    improves resource efficiency and latency predictability for tile-based ADS.
\end{abstract}

\bstctlcite{IEEEexample:BSTcontrol}

%%%%%% -- PAPER CONTENT STARTS-- %%%%%%%%
% \bstctlcite{IEEEexample:BSTcontrol}
% \input{tex/1_intro_xch.tex}
\section{Introduction}

Autonomous driving systems (ADS) are moving toward Level-4 (L4+) autonomy,
where the vehicle must handle open-road traffic without human
intervention~\cite{SAE-J-3016-2021, UNECE-grva-guidelines}.
For L4+ ADS, algorithms have moved from rule-based modular designs to
\ADSArch End-to-End (\ETEB) sensing-to-control pipelines that cover perception,
prediction, planning, and \aux monitoring. They also use redundant and diverse
DNN ensembles for fault resilience~\cite{NVsaftyAVReport2024}.
This puts dozens of heterogeneous DNNs on the chip at high update rates.
Each \ETEs chain must meet a strict latency constraint ($<$100\,ms), while
execution time still varies under dynamic driving conditions, by up to
3.3$\times$ in prior measurements~\cite{Gog2022-Eurosys-D3}.
\revision[blue][R5-Q1]{Unlike avionics systems that can dedicate isolated hardware to each function,
mass-produced ADS cannot afford such over-provisioning. Dense on-chip
colocation is therefore a cost necessity, but it introduces inter-task resource
contention.}

Recent designs from \revision[blue][R4-Q5]{Tesla~\cite{DojoMicro2023}\footnote{\href{https://en.wikipedia.org/wiki/Tesla_Autopilot_hardware\#Hardware_5}{\uline{AI5/6}}: derived from the Dojo tile-based architecture.}, NVIDIA~\cite{NV-Thor}, and Tenstorrent~\cite{Tenstorrent-2022-ISSCC}}
organize on-chip resources into independently controlled compute modules, 
called \emph{tiles}, connected by programmable NoC
fabrics. Each tile provides dedicated processing resources and on-chip memory
and is assigned DRAM space, allowing hardware-level spatial partitioning for co-located DNNs. 
Beyond isolation, this architecture exposes a unique scheduling knob:
a DNN's execution time decreases as its \emph{degree of parallelism} (DoP)
increases, which is proportional to the number of allocated tiles~\cite{Nv_simba}. 
The scheduler can allocate more tiles to a task during a
variation spike and release underutilized tiles to other co-located pipelines when demand falls.
Using this knob, however, requires spatial reallocation, which involves migrating
weights and intermediate features across tiles, incurring hundreds of microseconds execution stall, 
one to two orders of magnitude more than a CPU context switch.

Two scheduling paradigms also concern both utilization and latency guarantee, but neither
fits scheduling ADS workflow on \hwname accelerators. 
Existing real-time schedulers~\cite{Casini2019, McLean2020-tttech_tablescheduling,
baker1989cyclic} enforce strict isolation by compiling fixed DoP assignments
and temporal budgets offline, achieving near-zero scheduling overhead.
Recent \hwname accelerator schedulers~\cite{Liu2022-VELTAIR, Kim2023MoCA,
Ghodrati2020-Planaria}, by contrast, adjust DoP greedily at runtime
to maximize utilization.

Their underlying assumptions, however, do not hold in tile-based ADS. 
Static schedulers assume bounded worst-case execution time (WCET) and fix DoP accordingly, over-provisioning
spatial resources; \revision[blue][R1-Q1]{dynamic reservation servers (CBS~\cite{Lelli2016},
IRIS~\cite{marzario2002iris}, GRUB~\cite{lipari2000grub}) reclaim unused time
budgets but they keep a fixed DoP and cannot utilize spatial tunability.}
\revision[blue][R3-Q2]{Work-conserving schedulers assume reallocation as cheap.
In L4+ ADS, however, DNNs run from 10 to 240\,Hz~\cite{NVsaftyAVReport2024},
and each task event can trigger chip-wide reallocation. The accumulated
scheduling-induced overhead greatly affects \ETEs latency.}
\revision[blue][R3-Q1]{The most critical mismatch is deadline structure. These schedulers 
assume that each task has an explicit, independent deadline, while \ETEB ADS sub-deadlines are
jointly derived from a single \ETEs deadline. When an upstream node in the DAG
finishes early, downstream nodes that incur extra execution time
can use the slack to avoid an \ETEs deadline violation. A scheduler that ignores this DAG structure sees only local
slack or local deadline pressure, so it tends either to reserve too much or to
reschedule too late.}

We present \PaperName, a scheduling framework tailored for L4+ ADS on \hwname
accelerators, 
which can utilize DoP tunability in colocation with bounded reallocation cost.
\revision[blue][R3-Q1]{Unlike cluster schedulers that minimize cache-migration cost across homogeneous
cores, or hierarchical schedulers that bound memory and time-slice contention,
\PaperName targets the stop-migrate-restart penalty of spatial DoP reallocation,
using hardware-native isolation to confine its impact.}
\revision[blue][R6-Q1]{\PaperName's runtime scheduler is DAG-aware: it adjusts per-task DoP
within each partition and shares slack across DAG edges. The scheduler relies on
two mechanisms to keep those adjustments bounded.}
The \emph{elastic reservation} layer constrains \emph{when} DoP reallocation should
be triggered, using predicted task start and finish times.
The \emph{configurable isolation} bounds \emph{where} reallocations may
propagate by confining DoP changes within hardware-partitioned groups.
\revision[blue][R6-Q1]{Together, the scheduler absorbs short spikes and recycles underutilized tiles without frequent chip-wide reallocation.}

This paper makes three contributions:
\begin{itemize}[leftmargin=1.5em, labelsep=0.5em, nosep]
    \item We identify the fundamental mismatch between prior scheduling
    assumptions and tile-based ADS, formally modeling the spatio-temporal
    scheduling space that couples
    exploiting DoP tunability and controlling spatial reallocation overhead
    across DAG-structured \ETEs workflows under runtime variation
    (\cref{sec::existing_scheduling_policies}).
    \item We present \PaperName, whose DAG-aware runtime scheduler combines
    elastic reservation and configurable isolation into an 
    isolation-sharing space, exploiting DoP tunability at controlled reallocation
    cost without a binary choice between isolation and sharing
    (\cref{sec::ADS-Tile}).
    \item We implement and evaluate \PaperName on an industry- and
    academia-derived benchmark~(\cref{sec::eval}). \revision[blue][R5-Q2]{In
    deadline-critical cases, \PaperName uses up to \textbf{32\%} fewer tiles than
    work-conserving baseline schedulers while reducing reallocation-induced
    wasted processing capacity from \textbf{17\%}--\textbf{44\%} down to below
    \textbf{1.2\%}.}
\end{itemize}

\textbf{Organization.}
Section~\ref{sec::background} provides background on ADS workloads,
\hwname architectures, and the scheduling problem formulation.
Section~\ref{sec::existing_scheduling_policies} surveys representative scheduling paradigms, 
introduces the Guided Hybrid Allocation (GHA) framework for adapting them to \hwname ADS, 
and presents a case study that exposes their limits.
Section~\ref{sec::ADS-Tile} details the \PaperName spatio-temporal
isolation-sharing space and DAG-aware runtime scheduler.
Section~\ref{sec::eval} evaluates \PaperName against baselines.
Section~\ref{sec::related_work} surveys related work, and
Section~\ref{sec::conclusion} concludes.

% Section II: Background & System Model
% (2_bg_asplos26.tex internally \input's system_model.tex and problem_formulation.tex)
% \input{tex/2_bg_new_reduced.tex}
\section{Background and Problem Formulation}\label{sec::background}
This section defines the representative \ADSArch workflow, the \hwname platform model, and the scheduling abstraction used by \PaperName.

\subsection{Representative ADS Workflow}%
\label{sec::representative_ads_workflow}

% \textbf{Components.}
\begin{figure*}
  \centering
      %  \begin{overpic}[width=0.8\linewidth]{figures/pdf/ADS_ppt_brief.pdf}
      %   % \put(100,4){\includegraphics[width=0.26\linewidth]{figures/pdf/task_graph_abs.drawio.pdf}}
      %  \end{overpic}
    \begin{minipage}[c]{\linewidth}
        \centering
        \includegraphics[width=0.85\linewidth]{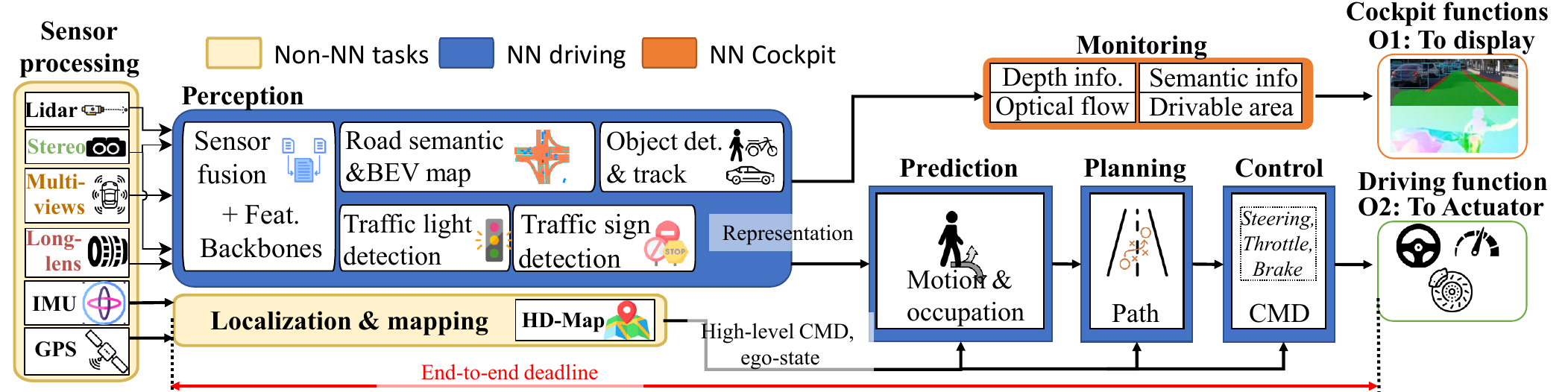}
        \caption{The workflow of a representative \ADSArch ADS benchmark.} \label{fig::e2e_workflow}
      \end{minipage}
  \vspace*{-0.5cm}
\end{figure*}
Figure~\ref{fig::e2e_workflow} depicts a typical \ADSArch \ETEs ADS architecture that consists of sensors and DNN modules
connected through a publish-subscribe message-passing model~\cite{valigi2021lessons}, forming an \ETEs workflow that delivers multiple end-to-end functions.
The pipelines start with sensing tasks that collect raw data from sensors and preprocess it. The formatted data are then delivered to DNN pipelines, which can be broadly categorized into two types based on their output targets~\cite{NVsaftyAVReport2024}:
\begin{enumerate}[leftmargin=0pt, itemindent=2.0em, labelsep=0.5em, label=(\roman*), nosep]
  \item (To the actuator, Figure~\ref{fig::e2e_workflow}, blue box) The \highlight{Driving function} generates vehicle control through five stages: perception, localization, prediction, planning, and control. The \textit{perception} stage fuses camera and LiDAR data~\cite{Yin2021-centerpoint} to build a representation of the environment used for detecting pedestrians, vehicles, traffic signals, and lanes.
  %\cite{Gan2021-Eudoxus}
  The \textit{Localization} stage detects the positions of objects and the vehicle, and informs the \textit{prediction} stage~\cite{Li2022-BEVFormer} to forecast their future movement. The \textit{Planning} stage computes waypoints that ensure safe and comfortable driving~\cite{Chen2022-LAV, hu2023UniAD}. The \textit{Control} stage finally translates these waypoints into steering and acceleration commands~\cite{Chen2022-LAV, Codevilla2018-ConditionalIL}.

  \item (To the cockpit, Figure~\ref{fig::e2e_workflow}, orange box) The \highlight{Digital cockpit functions} 
  % , referred to as \Aux pipelines in this paper, 
  enhance safety and comfort and are continuously expanding in scope and diversity~\cite{NVsaftyAVReport2024}.
  Capturing this trend, the workflow includes four \aux monitoring modules---road semantics, depth, dynamic targets, and optical flow---whose outputs are displayed on the cockpit screen to help drivers understand the vehicle's behavior.
\end{enumerate}

\subsection{The Evolution of ADS Computing Platforms}
\label{subsec::ads_platform_evolution}
% \begin{figure}[htbp]
%   \includegraphics[width=\linewidth]{figures/pdf/demo_spatial_dim_holistic.drawio.pdf}%270
%   \caption[short]{
%       (a) The architecture of \hwname accelerator, and spatial partitioning ability;
%       (b) Dynamic scheduling for colocation within a partition, where B is dequeued on completion, A scales up, and B is not influenced;
%       (c) using dataflow compiler to optimize data reuse of single DNN across multiple tiles.}
%       \label{fig::soc_overview}
%       \vspace{-0.5em}
% \end{figure}

\begin{figure}[htbp]
  \includegraphics[width=\linewidth]{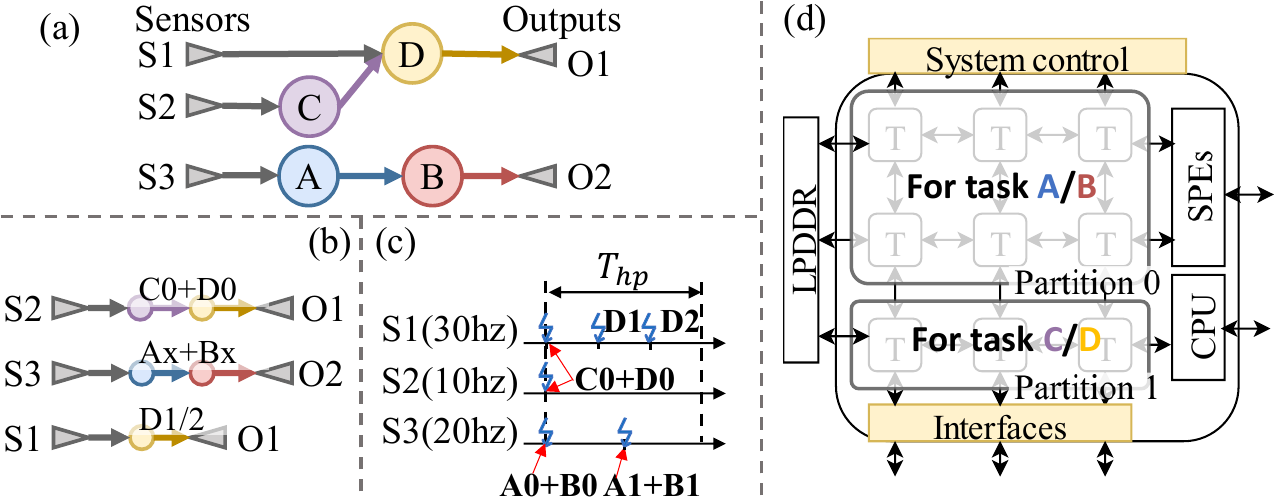}
  \caption[short]{
    (a) A simplified multi-rate workflow graph, (b) patterns of \ETEs task chains, (c) trigger offset of each pattern over $T_{hp}$ frames,
    (d) The architecture of the \hwname accelerator, and spatial partitioning ability.}
      \label{fig::soc_overview}
      \label{fig::exe_pattern}
      \vspace{-0.5em}
\end{figure}

Early ADS used tens of distributed Electronic Control Units~\cite{NV-Thor}, each dedicated to a single function, providing good isolation but with prohibitive cost for \ADSArch workflows.
Some accelerators~\cite{Kao2022-MAGMA,sun2024jigsaw} support colocation of multiple DNNs, but lack isolation between shared resources, causing resource contention~\cite{Kim2023MoCA}.
Next-generation \hwname platforms bridge both gaps: as shown in Figure~\ref{fig::soc_overview}d, a single SoC integrates DNN-computation tiles interconnected via a 2D-mesh NoC, sensor processing engines (SPEs), and a host CPU, supporting both multi-DNN colocation and hardware-enforced spatial isolation.
Crucially, the tile count allocated to a DNN is a runtime-adjustable degree of freedom, enabling the scheduler to trade spatial resources for latency---a key property we formalize in Section~\ref{sec::hw_model}.
These capabilities together motivate the formal system model we develop next.

\textbf{Timing criticality.}
Unlike hard real-time systems such as aircraft controls or vehicle chassis controls~\cite{Luo2019-UCB_safeos}, latency is not the only concern in ADS, and minimizing the risk of timeouts is a reasonable goal for two reasons:
(a) ADS is not the last line of defense for safety. A dedicated failure mitigation function (not an ADS function) serves as a safety backup to bring an ADS-equipped vehicle to a controlled stop in its path~\cite{SAE-J-3016-2021}.
(b) The main challenge of L4+ ADS is to address long-tail scenarios. Under limited hardware budgets, allocating resources for algorithmic redundancy is more practical than provisioning enough hardware to prevent all timeouts.
Therefore, in this paper, we specify timing as an \ETEs latency constraint with a probability guarantee that upper-bounds timeout risk, similar to firm real-time systems.

% \ETEs ADS is not the last line of defense safety,
% failure migration function (not an ADS function) performs as a safety backup to bring an ADS-equipped vehicle to a controlled stop in path~\cite{SAE-J-3016-2021}.
% 鍜岀‖瀹炴椂绯荤粺涓嶅悓
% 1. 鐢变簬浠诲姟娴佽礋杞界殑鍔ㄦ€佹€э紝绠楀姏闇€姹傜殑鏂瑰樊闈炲父澶э紝鍋氬埌鐧惧垎鐧句笉瓒呮椂闇€瑕佺殑绠楀姏鍐椾綑闈炲父楂樸€?
% 2. 褰撳墠宸ヤ笟瀹炶返鍊惧悜浜庯紝鐢ㄦ洿澶氱殑绠楀姏鍘绘敮鎸佺畻娉曠殑鍐椾綑锛岃€屼笉鏄笉瓒呮椂銆?
% 3. 鍦ㄥ啑浣欑殑绠楁硶璁捐涓嬶紝鍗曠偣瓒呮椂骞朵笉涓€瀹氬鑷寸郴缁熷け鏁堛€?
% 4. 绔埌绔殑ADS骞朵笉鏄畨鍏ㄦ渶鍚庨槻绾匡紝澶辫触杩佺Щ鍔熻兘锛堜笉鏄疉DS鍔熻兘锛夎捣鍒板浠戒綔鐢紝鍦ㄧ畻娉曞け鏁堢殑鏃跺€?performs simple maneuvers (e.g., braking or pulling over) to ensure a minimal risk condition~\cite{SAE-J-3016-2021}銆?
% 鍦ㄦ鑳屾櫙涓嬶紝鎴戜滑鎬绘槸璁や负缁欏畾鐨勭‖浠惰祫婧愬拰浠诲姟娴佹€绘槸鍖归厤鐨勶紝鎴戜滑鍙渶缁欏嚭鏈€灏忓寲瓒呮椂椋庨櫓鐨勮皟搴︽柟妗堬紝鎴栬€呬綔涓烘姌涓紝缁欏畾涓€涓疆淇″害锛屽鏋滃瓨鍦紝鎵惧埌瓒呮椂椋庨櫓涓嶈秴杩囬鍒剁殑鏂规銆?

%% system_model.tex
%% Subsections: Hardware Architecture Model (III.A.1) + Workload Model (III.A.2)
%% Responds to:
%%   Reviewer 1 Comment 2 (missing formal system model and notation table)
%%   Reviewer 1 Comment 3 (hyperperiod validity for data-driven DNN tasks)
%%   Reviewer 2 (DRAM channel separation, NUMA topology question)
%%   Reviewer 3 (memory bandwidth contention modeling, NUMA architecture)
%%
%% NOTE: This file is \input'd from tex/2_bg_asplos26.tex, replacing system_model_v0.tex.
%% The figure fig::exe_pattern / fig::soc_overview is defined earlier in 2_bg_asplos26.tex;
%% we reference it with \ref rather than redefining it here.

\subsection{System Model}\label{sec::system_model}

\revision[blue][R1-Q2]{This section formalizes the hardware architecture, the \ETEs ADS workload structure,
and the runtime variation factors. The probabilistic latency model and the notation table
are presented in Section~\ref{sec::latency_model}.}

% ─────────────────────────────────────────────────────────────────────
\subsubsection{Hardware Architecture Model}\label{sec::hw_model}

We model the \hwname accelerator as a heterogeneous SoC containing a total of $M$
identical DNN-computation \emph{tiles} alongside dedicated Sensor Processing Engines
(SPEs) for sensor preprocessing and a host CPU for system management.
Each tile integrates local computational units, private L1/L2 SRAM, and a lightweight RISC-V
control core; tiles communicate via a 2D-mesh Network-on-Chip (NoC), and a set of DRAM
memory controllers (MCs) are placed at the mesh boundary nodes.

\noindent\textbf{Spatial partitioning.}
The $M$ tiles can be carved into $S$ isolated \emph{partitions} (also called \emph{bins})~\cite{Nv_simba}.
Partition $s \in \{0, \ldots, S{-}1\}$ holds $|B_s|$ tiles, with $\sum_{s} |B_s| \leq M$.
Tiles belonging to different partitions do not share any resources besides the NoC,
structurally eliminating compute and cache interference.
As shown in Figure~\ref{fig::soc_overview}d, partitions also carry independent task
queues and RISC-V schedulers, enabling distributed per-partition runtime control
(Section~\ref{sec::reservation_runtime}).

\revision[blue][R1-Q4c]{Within a partition, task $v$ may be allocated $c_v$ tiles drawn from a finite
pre-compiled set $c_v^{\text{compiled}}$; this set is determined by dataflow search (Section~\ref{sec::multi_version_impl}).
The \emph{degree of parallelism} (DoP) of task $v$ is therefore proportional to $c_v$:
execution time scales approximately as $1/c_v$ (modulo memory-bound ceilings and NoC
communication overhead), giving the scheduler a continuous spatio-temporal
trade-off knob.}

\revision[blue][R3-Q3,R2-Q1,R4-Q2]{\noindent\textbf{Memory and interconnect model.}
Memory controllers (MCs) are distributed at the 2D-mesh boundary nodes, and each partition is
statically bound to its nearest MC, establishing partition-local memory affinity.
Unlike conventional NUMA where any node may access any remote memory region,
cross-partition memory traffic is prohibited by design: each partition addresses only the
data served by its bound MC(s).
Because ADS data packets carry sensor timestamps and task-priority metadata,
we assume MCs support a \emph{priority-aware request queue} to provide temporal isolation
among co-located tasks---a lighter-weight approach than MoCA-style centralized bandwidth throttling~\cite{Kim2023MoCA}, better suited for real-time systems.}

\revision[blue][R3-Q3,R2-Q1,R4-Q2]{On the NoC side, the two traffic classes have distinct contention profiles.
\emph{Intra-partition compute communication} (collective operations among co-located tiles)
stays within the contiguous tiles in a rectangular partition and does not cross partition boundaries.
\emph{Tile-to-MC memory traffic}, in contrast, traverses the mesh and introduces latency uncertainty.
To alleviate cross-partition NoC contention, fixed paths can be established between each partition and its bound MCs.
With these fixed paths, the average hop count can be bounded by a constant, and the main source of variation is queuing latency on the path.
These factors are captured by the probabilistic latency model in Section~\ref{sec::latency_model}.
}

% ─────────────────────────────────────────────────────────────────────
\subsubsection{Workload Model}\label{sec::workload_model}
An ADS workflow is modeled as a Directed Acyclic Graph (DAG) $G(V, E)$, where
$V = V_{\text{sen}} \cup V_{\text{dnn}}$ and $E$ encodes data-flow dependencies.
Tasks in $V_{\text{sen}}$ are sensor preprocessing modules; tasks in $V_{\text{dnn}}$
are DNN inference modules.
A task is \emph{activated} when it is ready to execute; a single run is called a \emph{job}.
An \concept{end-to-end chain} is a sequence of tasks from a sensor source to an
actuator or display sink.

% \noindent\textbf{Activation model and hyperperiod.}
\revision[blue][R1-Q3]{Sensor tasks are activated by external hardware timers at strictly periodic intervals;
DNN tasks are \emph{data-driven}: task $v \in V_{\text{dnn}}$ becomes ready only when
all its predecessors complete their current job.
% Here, ``data-driven'' refers to \emph{precedence constraints} between tasks, not arbitrary
% or aperiodic data arrivals.
Since all input data originates from periodic sensors, the \emph{data-dependency patterns}
of each task instance repeat over the \concept{hyper-period}
$T_{hp} = \text{lcm}\{T_v \mid v \in V_{\text{sen}}\}$.
By unrolling the DAG over $T_{hp}$, each task $v$ decomposes into $N_v = T_{hp}/T_v$
instances (e.g., A0, A1, B0, B1 in Figure~\ref{fig::exe_pattern}b--c) with a static,
deterministic dependency structure, making standard hyperperiod scheduling analysis
fully applicable.}

% \noindent\textbf{End-to-end timing constraints.}
An \ETEs path $\xi$ must complete within an \concept{end-to-end latency constraint}
$\mathcal{D}_{e2e}$, measured from the activation of the source sensor to the completion
of the sink task.
For distributed per-task scheduling, $\mathcal{D}_{e2e}$ is decomposed into individual
\concept{latency budgets} $l_v$, which establish \concept{sub-deadlines}
$ddl_{\text{sub},v} = t_v + l_v$, where $t_v$ is the planned start offset of
task $v$.
The decomposition satisfies $\sum_{v \in \xi} l_v \leq \mathcal{D}_{e2e}$ by
construction; the offline Phase-I solver (Section~\ref{sec::GHA}) determines the
budget assignment $l_v$ jointly with the spatial allocation $c_v$.

% ─────────────────────────────────────────────────────────────────────
\subsubsection{Probabilistic Latency Model}\label{sec::latency_model}

To formally evaluate the \ETEs latency guarantee of a workflow under variation,
we adopt a probabilistic model. Specifically, we introduce two random variables to
capture the execution-latency randomness caused by two sources of variation:
\begin{itemize}[leftmargin=*, nosep]
  \item \textbf{F1: Execution variation.}
  For DNN tasks, execution time fluctuates with input complexity (e.g., the number of
  detected objects in a scene) and scenario-driven model-size switching; the
  99th-percentile execution time can exceed the mean by 3.3$\times$~\cite{Gog2022-Eurosys-D3}.
  For sensing tasks, variation arises from hardware nondeterminism (e.g., cache effects)
  and software preprocessing overheads~\cite{valigi2021lessons}.
  \item \textbf{F2: Inter-task interference.}
  When co-location is enabled within a partition, multiple tasks contend for DRAM bandwidth.
  Concurrent I/O requests queue at the memory controller, introducing I/O latency variation.
\end{itemize}

Let $W_v$ denote the random variable for the arithmetic workload of task $v$
(arising from F1), whose distribution is obtained by profiling traffic conditions
from open-road data or digital twin simulations~\cite{NVsaftyAVReport2024}.
Let $I_v$ denote the random variable for the I/O latency of task $v$ (arising from F2).
It contains two components: a constant latency component determined by the average
tile-to-MC hop count, and a stochastic queuing latency component.
Following prior work~\cite{Chang2005}, we model the queuing component with an M/M/1 model,
which yields a \emph{shifted exponential distribution} whose tail grows with DRAM utilization.

Given $c_v$ tiles allocated to task $v$, the compute component scales as $W_v / (c_v \cdot P)$,
where $P$ is the processing power (FLOPS/cycle) of a single tile.
Assuming worst-case correlation between workload and I/O contention,
the actual execution latency is bounded by:
\begin{IEEEeqnarray}{rCl}
  L_v(q, c_v) \triangleq \frac{W_v^{(q)}}{c_v \cdot P} + I_v^{(q)}
  \label{eq::latency_model}
\end{IEEEeqnarray}
where $W_v^{(q)}$ and $I_v^{(q)}$ are the $q$-th quantiles of their respective
distributions.
By construction, $\Pr[L_v(c_v) \leq L_v(q, c_v)] \geq q$: task $v$ completes within
$L_v(q, c_v)$ with probability at least $q$.
Equation~\ref{eq::latency_model} therefore gives an independent per-task probabilistic latency bound.

For sensor preprocessing tasks, each source has a dedicated SPE and I/O port,
so we model their tail latency directly as a univariate distribution
$L_v(q) = \mathcal{D}_v^{(q)}$.

\revision[blue][R1-Q1]{\noindent\textbf{Role of quantile $q$.}
The parameter $q$ serves as a \emph{joint spatio-temporal knob}: a higher $q$ requires a
larger tile allocation $c_v$ (spatial resource) and a larger latency budget $l_v$
(temporal resource) to absorb the longer tail of variation.
This extends the classical probabilistic scheduling guarantee---previously applied only
to the temporal dimension in reservation servers such as CBS~\cite{Lelli2016} and
GRUB~\cite{lipari2000grub}---to the spatial dimension ($c_v$) of \hwname accelerators.}

\noindent\textbf{Scope note.}
% Writing note: Avoid stating this as "quantile sub-additivity."
% Quantiles are not sub-additive in general, and "not perfectly correlated" is
% not a sufficient condition for Q_q(X+Y) <= Q_q(X)+Q_q(Y). The text below
% frames summed per-task tail-latency budgets as a profile-observed conservative envelope.
Across the latency profiles used in this paper, directly summing per-task tail-latency budgets typically overestimates the observed E2E tail latency, because tail events from different tasks rarely align within the same chain instance.
We instantiate this conservative provisioning envelope as $\sum_v L_v(q, c_v)$ and refer to the resulting profile-observed margin as \emph{tail-composition headroom}.
This empirical conservatism is quantified in Section~\ref{subsec:limitations}
and is systematically recovered by the runtime slack-sharing mechanism
(Section~\ref{sec::hybrid-grained_isolation}).

\subsubsection{Notation Table}\label{sec::notation_table}

\revision[blue][R1-Q2]{Table~\ref{tab::notation_table} summarizes the main symbols used throughout the paper.}
With this notation in place, Section~\ref{sec::problem_formulation} next frames the scheduling
task as a spatio-temporal bin-packing problem and shows how prior schedulers map to special
cases of this formulation.
\begin{table}[ht]
    \centering
    \caption{Key notation for spatio-temporal scheduling}
    \label{tab::notation_table}
    \resizebox{\linewidth}{!}{%
    \begin{tabular}{c l}
        \toprule
        \textbf{Symbols} & \textbf{Meaning} \\
        \midrule
        $G(V,E)$, $\xi$, $v$ & Workflow DAG, E2E chain/path, and task node \\
        $\mathcal{D}_{e2e}$, $ddl_{\text{e2e}}$ & E2E latency constraint and chain-instance deadline \\
        $q$, $T_{hp}$ & Target quantile and hyper-period \\
        $M$, $S$ & Total tiles and partitions \\
        $B_s$, $x_{vs}$ & Partition capacity $|B_s|$ and task-to-partition mapping \\
        $c_v$, $c_v^{(t)}$ & Offline tile allocation and runtime tile quota \\
        $l_v$, $t_v$ & Task latency budget and planned start offset \\
        $ddl_{\text{sub},v}$ & Task sub-deadline \\
        $L_v(q,c_v)$ & Per-task probabilistic latency bound \\
        \bottomrule
    \end{tabular}
    }
\end{table}

%% problem_formulation.tex
%% Section III-D: Problem Formulation — Spatio-Temporal Scheduling Space
%% Extracted from 0-motivation-zyp.tex L93-131, restructured per sec2-4_COT_v3.md
%% NOTE: This file is \input'd from tex/2_bg_asplos26.tex, after system_model.tex.

\subsubsection{Problem Formulation}\label{sec::problem_formulation}

Prior \CyclicName-like (time-reservation) and \TpFirstName-like (throughput-driven colocation) schedulers optimize either the temporal or the spatial dimension in isolation.
In contrast, an ADS on a \hwname accelerator must jointly decide \emph{when} (start time and duration) and \emph{where} (logical/physical mapping) to execute tasks in a workflow graph $G(V, E)$ to satisfy $\mathcal{D}_{e2e}$ while maximizing utilization. This joint spatio-temporal problem is computationally intractable due to the tight coupling among (i) execution order and concurrency on the DAG, (ii) temporal assignment (time), and (iii) spatial assignment (bins/tiles).

To formally define the scheduling space, we frame the problem as a spatio-temporal bin-packing problem:
$|V|$ tasks in the workflow graph are mapped to $M$ tiles, where tasks and tiles are, respectively, the items to be packed and the bins in which they are placed.
$M$ tiles are divided into $S$ partitions, each viewed as a \concept{bin} with shape $(|B_s|, T_{hp})$ (Eq.~\ref{eq:space_bins}).
Tasks are the \textbf{items} to be packed, with spatial size ($c_v$ tiles) and temporal size (latency budget $l_v$) (Eq.~\ref{eq:space_items}).
\revision[blue][R1-Q4a]{The goal of a static scheduler is to produce a feasible packing plan.
Spatially, each item is mapped to a bin ($x_{vs} \in \{0,1\}$); temporally, each item is assigned an offset ($t_v$), where $ddl_{\text{sub},v} = t_v + l_v$ is the sub-deadline (Eq.~\ref{eq:space_map}).
\begin{compacteqn}
  \begin{IEEEeqnarray}{s'cl}
    \IEEEyesnumber\IEEEyessubnumber*
    Bin partitioning: & (|B_1|, \ldots, |B_S|) \times T_{hp} & \; \text{(capacity)} \label{eq:space_bins}\\
    Items: & v \mapsto (c_v, l_v) &\; \text{(shape)} \label{eq:space_items}\\
    Mapping: & v \mapsto (x_{vs}, t_v) &\; \text{(position)}  \label{eq:space_map}
  \end{IEEEeqnarray}
  \end{compacteqn}}
This unified view allows us to express prior schedulers as special cases:
{\TpFirstName-like} schedulers determine the temporal scheduling on-the-fly and only decide spatial packing ($c_v$) on a single large bin ($S=1, |B_1|=M$).
{\CyclicName-like} schedulers only search a mapping scheme offline, assuming a fixed partition scheme, and each task has a predetermined WCET under fixed spatial size.
\highlightit{Note that}, unlike vanilla bin-packing, the shapes of both items and bins are not predetermined, and the packing result must meet the \ETEs latency constraints under runtime variation.

% Section IV: Adapting Existing Schedulers & Motivation
%% baseline_paradigms.tex
%% Section IV: Adapting Existing Schedulers & Motivation
%% Section IV-A: Representative Scheduling Paradigms
%% Extracted from 0-motivation-zyp.tex L1-67, restructured per sec2-4_COT_v3.md

\section{Adapting Existing Schedulers \& Motivation}
\label{sec::existing_scheduling_policies}

Having formalized the scheduling space in Section~\ref{sec::problem_formulation},
we now survey two representative paradigms that can be adapted to solve this problem,
and examine their fundamental limitations on \hwname ADS.
No prior work has studied how to deploy ADS workloads on \hwname accelerators.
To bridge this gap, we first introduce two representative scheduling paradigms
and then develop a guided offline allocation framework to adapt them to the \hwname
ADS setting, enabling a fair experimental comparison.

\subsection{Representative Scheduling Paradigms} \label{subsec::baseline_idea}
Existing schedulers that can be adapted to \hwname ADS fall into two categories:
reservation-based and throughput-driven schedulers.
Both address latency and utilization under variation (F1--F2), yet each exploits only
part of the spatio-temporal scheduling space defined in Section~\ref{sec::problem_formulation}.

\begin{figure}[h]
  \centering
  \includegraphics[width=0.99\linewidth]{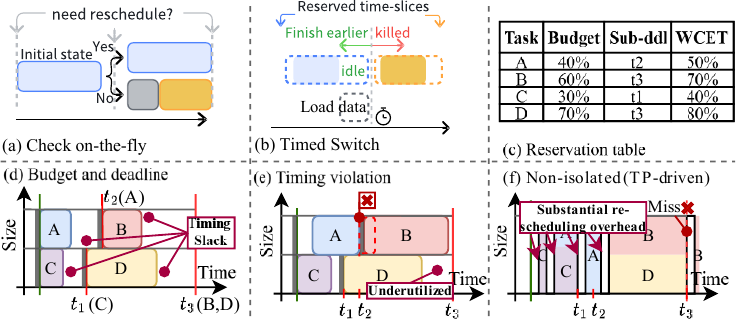}
  \caption{Basic idea of two representative schedulers: \CyclicName and \TpFirstName.}\label{fig::sched_space_plus_tab}
  \vspace{-1em}
\end{figure}

\subsubsection{Fully-isolated, time-multiplexing scheduler (\CyclicName)}\label{sec:full_isolated_scheduler}
Recent ADS schedulers such as D3~\cite{Gog2022-Eurosys-D3} and table-scheduling approaches~\cite{McLean2020-tttech_tablescheduling} manage time slices on multiprocessors via static reservations~\cite{Casini2019}.
The key idea is to reserve each task a fixed \execTime budget on a fixed core.
In the context of \hwname ADS, we can assign each task a sub-deadline, divide tiles into partitions, and reserve per-task budgets within each partition, recorded by a reservation table \revision[blue][R2-Q2]{(Figure~\ref{fig::sched_space_plus_tab}c)}. As shown in \revision[blue][R2-Q2]{Figure~\ref{fig::sched_space_plus_tab}d}, tiles are split into two partitions, with \singlefuncs-A/B and C/D mapped with reserved budgets.
 Any task that overruns its budget or misses a sub-deadline is terminated, preventing its overrun from delaying other tasks. For static reservation, preemption is forbidden and the scheduler is triggered only at predetermined time boundaries.
 A timed switch enforces slot boundaries \revision[blue][R2-Q2]{(Figure~\ref{fig::sched_space_plus_tab}b)}: within each period, a task may run only in its reserved slot on its pinned core and is terminated if it overruns the budget. \textbf{Strengths:} As resource bindings are fully static, rescheduling overhead is negligible, and latency is independent of system load.

\subsubsection{Non-isolated, colocation-aware scheduler (\TpFirstName)}%
\label{sec::non-isolated_spatial-aware_scheduler}
\revision[blue][R2-Q8]{Recent studies such as Planaria~\cite{Ghodrati2020-Planaria}, VELTAIR~\cite{Liu2022-VELTAIR}, and MoCA~\cite{Kim2023MoCA} have explored multi-task colocation on \hwname accelerators in cloud DNN serving scenarios.
We collectively refer to this family as \TpFirstName and use Planaria as its representative in our evaluation.}
The core idea is to maintain a deadline-driven task queue and trigger on-the-fly rescheduling whenever the queue changes \revision[blue][R2-Q2]{(Figure~\ref{fig::sched_space_plus_tab}a)}: all available tiles are redistributed among ready tasks to keep every tile saturated.
In the \hwname ADS context, activated tasks are treated as independent jobs, each with a deadline. As shown in \revision[blue][R2-Q2]{Figure~\ref{fig::sched_space_plus_tab}e and Figure~\ref{fig::sched_space_plus_tab}f}, initially all tiles serve \singlefunc-C; once \singlefunc-A becomes ready, tiles are reallocated at sub-\singlefunc granularity. Thin white boxes represent computational stalls introduced by rescheduling.
\textbf{Strengths:} Because resources are never left idle even under varying load, this approach achieves high utilization.

%% gha_offline.tex
%% Section IV-B: Guided Hybrid Allocation (GHA)
%% Assembled from 0-motivation-zyp.tex L56-67, L137-261 and two-phase.tex L40-132

\subsubsection{Adapting \CyclicName and \TpFirstName to tile-based ADS}

Neither \CyclicName nor \TpFirstName was designed for tile-based accelerators.
Adapting them to \hwname ADS requires solving three non-trivial problems:
\begin{enumerate*}[leftmargin=*, nosep, label=({\roman*})]
 \item First, ADS workflows provide only \ETEB latency constraints. However, both \CyclicName and \TpFirstName require per-task sub-deadlines. Consequently, we need to allocate the \ETEB latency budgets to individual tasks.
 \item Additionally, the \hwname architecture comprises hundreds of tiles where a single DNN can use multiple tiles. However, \CyclicName assumes one-to-one core-task bindings, and tile-level mapping is too fine-grained to be applied directly. Consequently, we need to abstract a coarse-grained multi-core execution model by allocating tiles into multiple partitions.
 \item Most importantly, ADS requires predictable latency guarantees under runtime variation, while none of these schedulers can provide a variation-aware performance model to guide the above two allocation processes.
\end{enumerate*}
To address these challenges, we propose the Guided Hybrid Allocation (GHA) compiler that serves as the common adaptation layer for any scheduling paradigm on \hwname ADS, refining the plan from an abstract ADS specification into a concrete logical schedule.
% \PaperName runtime is able to track the distance of available tiles use $MaxDist$ to compensate this effect,
% which can be modeled by randomly mapping tasks to different physical tiles at runtime.

\subsection{Guided Hybrid Allocation}\label{sec::GHA}
\revision[blue][R1-Q4b]{%
GHA decomposes the joint spatio-temporal scheduling problem
(Section~\ref{sec::problem_formulation}) into three successive phases:
\begin{itemize}[leftmargin=*, nosep]
  \item \textbf{Phase~I\,---\,Chain-by-Chain Slack Assignment.}
    Given the DAG topology, Phase~I determines the per-task shape $(c_v, l_v)$ on each E2E chain,
    minimizing peak tile usage subject to the end-to-end deadline $\mathcal{D}_{e2e}$.
  \item \textbf{Phase~II\,---\,Spatial Partitioning.}
    Given the per-task shape from Phase~I, Phase~II decides the task-to-partition mapping
    $x_{vs}$ and partition capacities $|B_s|$, minimizing total capacity while balancing
    utilization across partitions.
  \item \textbf{Phase~III\,---\,Intra-partition Temporal Compaction.}
    Given the partition layout from Phase~II, Phase~III refines the temporal placement
    $t_v$ within each partition and applies first-fit-decreasing repacking
    to enforce the total tile budget $\sum_s |B_s| \leq M$.
\end{itemize}
}
% This decomposition preserves feasibility under variation, exposes clear tuning knobs via $(q_A, q_B)$ for robustness–efficiency trade-offs, and yields an implementable schedule that combines static predictability with dynamic resilience.

\subsubsection{Compilation Prerequisite}
\label{sec::compilation_prereq}

\revision[blue][R1-Q4c]{GHA requires a pre-compiled set of DNN implementations with varying
tile counts ($c_v^{\text{compiled}}$) and latency profiles ($L_v(q, c_v)$),
generated by dataflow search (Section~\ref{sec::multi_version_impl}).
To support runtime switching among implementations, all versions share the same checkpoint positions at operator boundaries. The switching behavior is detailed in Section~\ref{sec::impl_switching}.
}

\begin{figure}
  \centering
  \includegraphics[width=0.99\linewidth]{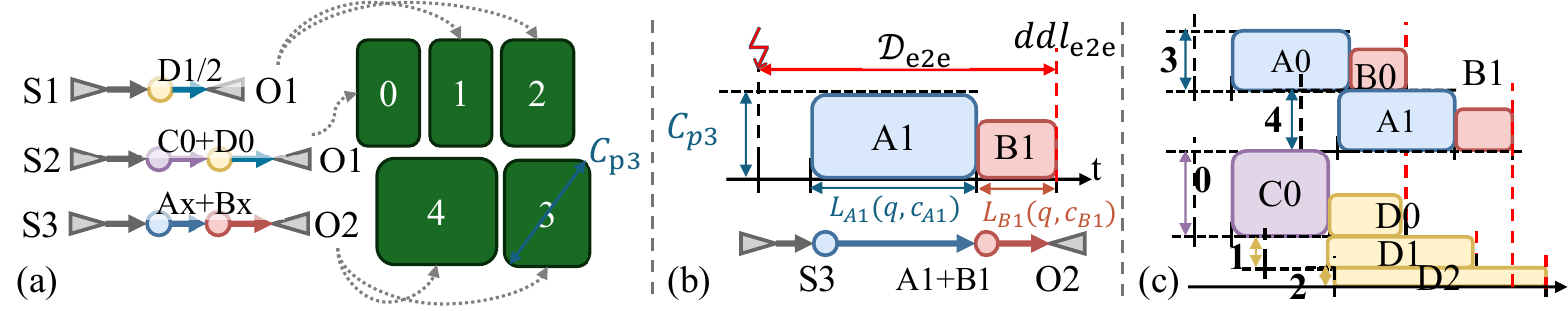}
  \caption{Guided Phase I for the scheduling problem.}\label{fig::guided_solver}
  \vspace{-1em}
\end{figure}

\subsubsection{Phase I: Chain-by-Chain Slack Assignment}\label{sec::Phase1}
\revision[blue][R2-Q3]{%
Figure~\ref{fig::guided_solver} illustrates the overall process of Phase I.%
}
To simplify position assignment, we adopt a chain-based decomposition strategy: each \ETEs chain is isolated into a separate partition, with tasks executing sequentially within that partition (\highlightuline{mirroring the \CyclicName model}) (Figure~\ref{fig::guided_solver}a). This reduces the full spatio-temporal problem to a series of simpler per-chain subproblems, where we determine the shape ($c_v, l_v$) for each task while keeping the execution order fixed \revision[blue][R2-Q3]{(Figure~\ref{fig::guided_solver}b)}.
% 最后一个任务的结束为止，早于端到端deadline，每个item的其实位置，不能早于上游节点对应的块。
% 每个item的shape有约束。1. 任务并行度的限制：For example, the perception DNNs generate large backbone networks, such as YOLOX and Transformer~\cite{Li2022-BEVFormer, Ge2021-YOLOX}, whereas the 
% planning and control DNNs are usually small models, such as MLP and GRU, thus varying greatly in the supported maximum parallel degree. 2. 设计规则的约束：只能从a list of configurations that is 
% compiled in advance (i.e.,\(c_v^{compiled}\)) 选择。
% 这个问题可以被抽象为一个二次规问题：
For each chain, the shape assignment problem can be formulated as a constrained optimization problem:
\begin{compacteqn}
\begin{IEEEeqnarray}{r'c'l}
\min & &\max_{v} c_v \IEEEyesnumber \label{eq:phase1_obj} \\
\text{s.t.} &\forall v  & s_{last} + l_{last} \leq \mathcal{D}_{e2e} \IEEEyesnumber \IEEEyessubnumber* \label{eq:phase1_e2e} \\
& &s_v \geq \max_{u \in \text{pred}(v)} (s_u + l_u), \label{eq:phase1_dep} \\
& &l_v \geq L_v(q, c_v),  \IEEEyesnumber \IEEEyessubnumber \label{eq:phase1_lat} \\
& &c_v \in \mathbb{Z}^+, \; c_v^{\min} \leq c_v \leq c_v^{\max} \text{ or } c_v \in c_v^{\text{compiled}}, \label{eq:phase1_core}
\end{IEEEeqnarray}
\end{compacteqn}
where $L_v(q, c_v)$ is the probabilistic latency bound defined in Eq.~\ref{eq::latency_model}.
\highlight{From a workflow-graph perspective}, Eq.~\ref{eq:phase1_obj} minimizes the peak tile usage across all tasks. Constraint Eq.~\ref{eq:phase1_e2e} ensures that the \ETEB deadline is met, while Eq.~\ref{eq:phase1_dep} enforces graph topological order (each item cannot start earlier than the completion of its predecessor blocks). Constraint Eq.~\ref{eq:phase1_lat} links the latency budget to resource allocation via the latency model. 
\revision[blue][R2-Q3]{Figure~\ref{fig::guided_solver}c illustrates the final result of Phase I.}

\highlight{From a per-task perspective}, Eq.~\ref{eq:phase1_core} captures practical constraints: (1) \emph{Parallelism limits} imposed by DNN architecture---for example, perception DNNs at the upstream of the chains (e.g., Transformer~\cite{Li2022-BEVFormer}) are inherently larger than models at the tail, supporting higher degrees of parallelism; and (2) \emph{Design-rule constraints} that restrict tile allocation to a pre-compiled list of valid configurations (\(c_v^{\text{compiled}}\)). 
This process is repeated chain-by-chain, ordered by their criticality and total load; previously assigned nodes keep their allocations and consume part of the remaining deadline on subsequent chains, as described in Algorithm~\ref{alg::multi_chain_slack_distribution}.

\begin{algorithm}
  \setstretch{0.95}
  \caption{Multi-Chain Slack Distribution (Chain-by-Chain)}
  \label{alg::multi_chain_slack_distribution}
  \footnotesize
  \begin{algorithmic}[1]
    \renewcommand{\algorithmicrequire}{\textbf{Input:}}
    \renewcommand{\algorithmicensure}{\textbf{Output:}}
    \Require Chains set $\{\xi\}$, E2E deadlines $\mathcal{D}^{(\xi)}_{e2e}$; workflow graph $G$; quantile $q$; Latency model $L_v(q, c_v)$
    \Ensure Set $\Omega=\{(c_v, l_v)\}$ for all assigned tasks
    \State $\Omega \leftarrow \varnothing$ \Comment{assigned nodes and their $(c,l)$}
    \State Sort $\{\xi\}$ by priority function
    \For{each chain $\xi \in \{\xi\}$}
      \State $C \leftarrow$ nodes of $\xi$ in topological order
      \Comment{Remove assigned nodes};
      \State $Done \leftarrow C\cap \Omega, U \leftarrow [\, C\setminus \Omega\,]$
      \State $D^{\text{rem}} \leftarrow \slack_{e2e} - \sum_{v\in Done} l_v$ \label{alg::rem_deadline}
      \If{$ U=\varnothing$} \textbf{continue};
      \EndIf
      \Comment{Assign for unassigned nodes};
      \State $(c_v,l_v)_{v\in U} \leftarrow \textsc{SolveSubChain}(q, G, U, D^{\text{rem}}, Done)$ \Comment{minimize peak tiles s.t. $\sum l_v\le D^{\text{rem}}$}
      \State $\Omega \leftarrow \Omega \cup \{(v, c_v, l_v)\}_{v\in U}$
    \EndFor
    \State Let $V_{topo}$ be the vertices of $G$ in topological order.
    \State Initialize $s_v \leftarrow 0, e_v \leftarrow 0$ for all $v \in V$.
    \For{each task $v$ in $V_{topo}$}
      \State $s_v \leftarrow \max(\{e_u \mid (u, v) \in E\})$ \Comment{Start after latest predecessor}
      \State $e_v \leftarrow s_v + l_v$ \Comment{End after its optimistic budget}
    \EndFor
  \end{algorithmic}
\end{algorithm}

Although Phase I provides an initialization for item shape, bin-capacity, and item-to-bin mapping,
it needs further refinement for three reasons: 1) initial packing result may be loose, which means poor utilization; 2) depending on the runtime scheduling policy, we need to tune the task-to-partition mapping to exploit the sharing opportunities; 3) although trying to minimize tile usage, maximum tile usage $M$ is not enforced.
These are addressed in Phase II and Phase III respectively.

\subsubsection{Phase II: Spatial Partitioning}\label{sec::Phase2}
This phase compacts the packing layout in the space, by clustering tasks into multiple partitions, 
which can be formulated as a 2D multiple packing problem, with horizontal position fixed.
We split all overlapping task execution intervals into a set of disjoint windows $T$, 
where $A_t$ is the tasks active in window $t$, and $d_t$ is the window duration. 
% \begin{compacteqn}
% \begin{IEEEeqnarray}{r'l's}
%   \min & w_1 \cdot \sum\nolimits_{s=0}^{S-1} |B_s|
%         - w_2 \cdot \text{Score}_{\text{affinity}}
%         + w_3 \cdot \text{Score}_{\text{balance}} & \nonumber \\
%   \text{s.t.} 
%   & \sum\nolimits_{s=0}^{S-1} x_{vs} = 1, 
%     \quad \forall v \in V
%     & \label{eq:phase2_assignment} \IEEEyesnumber \IEEEyessubnumber* \\
%   & \sum\nolimits_{v \in A_t} c_v x_{vs} \leq |B_s|,
%     \quad \forall s \in [0,S),\; \forall t \in T
%     & \label{eq:phase2_capacity}
% \end{IEEEeqnarray}
% \end{compacteqn}
\begin{compacteqn}
  \begin{IEEEeqnarray}{r'l's}
    % & \forall s \in [0, S), \forall v \in V, t \in T &\IEEEeqnarraynumspace \\
    % & \sum\nolimits_{s=0}^{S-1} x_{vs} = 1,  (\text{One bin per task}) \label{eq:phase2_assignment} \IEEEyesnumber \IEEEyessubnumber* \\
    % & \sum\nolimits_{v \in A_t} c_v x_{vs} \leq |B_s|. (\text{Capacity}) \label{eq:phase2_capacity} \\
    & \sum\nolimits_{s=0}^{S-1} x_{vs} = 1, 
    \quad \forall v \in V (\text{One bin per task})
    &  \label{eq:phase2_assignment} \IEEEyesnumber \IEEEyessubnumber* \\
  & \sum\nolimits_{v \in A_t} c_v x_{vs} \leq |B_s|,
    \quad \forall s \in [0,S),\; \forall t \in T (\text{Capacity})
    & \label{eq:phase2_capacity}  
    \end{IEEEeqnarray}
  \end{compacteqn}

% This phase compacts the packing layout in the spatial dimension by searching for the number of logical partitions and the corresponding item-to-bin mapping.
% Let $s \in [0, S)$ denote a partition index and $x_{vs} \in \{0, 1\}$ be a binary variable indicating whether task $v$ is mapped to partition $s$.
% The objective is to find $\{x_{vs}, \forall v \in V, s \in [0, S)\}$ given $S$. 
% The search process is guided by finding a mapping that minimizes the total bin sizes ($\sum |B_m|$), maximizes the total affinity score ($A$), and balances utilization across bins. 
% The formulation is as follows:
% by three criteria: 1) total bin size 2) task affinity 3) utilization balance: 
For a candidate bin count $S$, the optimizer selects the task-to-bin
mapping $x_{vs}$ and the induced bin capacities $|B_s|$ by trading off three
criteria: spatial compactness, affinity-preserving colocation, and load
balance across bins.
\begin{compacteqn}
  \begin{IEEEeqnarray}{r'l'l}
    \IEEEyesnumber \IEEEyessubnumber*
    \min & w_1 \cdot \sum\nolimits_{s} |B_s| - w_2 \cdot \text{Score}_{\text{affinity}} + w_3 \cdot \text{Score}_{\text{balance}} \\
    & \text{Score}_{\text{affinity}} = \sum\nolimits_{v,k,s} \beta_{vk} x_{vs} x_{ks} & \label{eq:phase2_affinity} \\
    &U_s =
    {
    \sum\nolimits_{t \in T}
    (\sum\nolimits_{v \in A_t} c_v x_{vs}) d_t
    }/{
    (|B_s| \cdot T_{hp})
    } &  \label{eq:phase2_utilization} \\
    &\text{Score}_{\text{balance}} = \max_{s \in [0, S)} U_s - \min_{s \in [0, S)} U_s & \label{eq:phase2_balance}
  \end{IEEEeqnarray}
\end{compacteqn}
Affinity defines the relatedness of tasks; for example, each task is preferred to be placed in the same bin as its predecessors and successors. Although at this phase we still assume mapping to logical tiles, tasks mapped to the same bin are more likely to be mapped to the same or adjacent tiles in the physical graph, bringing lower communication latency.
The total affinity score $Score_{\text{affinity}}$ is then formulated as the number of edges that do not cross bins (Eq.~\ref{eq:phase2_affinity}).
The utilization score computes each bin's hyper-period utilization
(Eq.~\ref{eq:phase2_utilization}) and penalizes imbalance across bins
(Eq.~\ref{eq:phase2_balance}).
\begin{figure}
  \centering
  \includegraphics[width=0.9\linewidth]{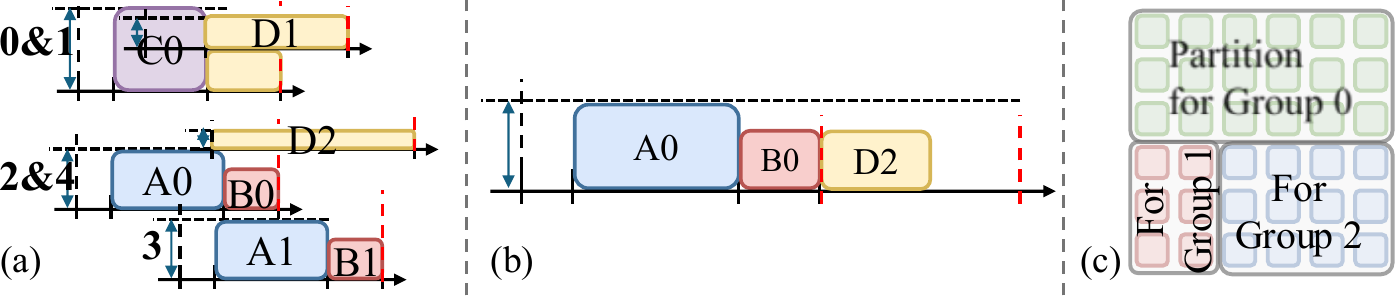}
  \caption{Guided phases II--III for the scheduling problem.}\label{fig::guided_solver2}
  \vspace{-1em}
\end{figure}
As shown in Figure~\ref{fig::guided_solver2}a, bins 0 and 1 are combined because of data affinity, and bins 2 and 4 are combined to improve bin load balance. However, bin coalescing cannot eliminate empty time slices (Bin \textbf{2\&4}) because of the task shape mismatch.
At this point, we obtain all scheduling information needed by \TpFirstName, which is specified with one partition ($S=1$), and by \CyclicName, which is specified using fewer bins with no temporal overlap in each bin.

\subsubsection{Phase III: Temporal Compaction}\label{sec::Phase3}
% 在前面两个阶段，虽然都将tile 数量作为优化目标，但是，由于保守的初始化，整体packing 结果可能比较松散。
% 在这个阶段，我们约束tile总量不超过M。为了达到这一目标，首先，我们按照比例，重分配每个bin的tile数量。
% 然后，在每个fixes the spatial partitioning from Phase 2,
% independently refines the 每个箱内，item的shape和temporal position，能够得到item更加精细的分配。
% allocation of tasks in each partition, which respectively is a bin-packing problem.
% 有很多可用开源的装箱工具，我们选择了一种基于First fit方法的启发的方法，解决这个问题，经过多轮迭代，返回每个bin内，物体的size和位置。
This phase enforces total tile usage $\leq M$ by compacting the packing result along the time dimension while maintaining the spatial assignment from Phase II.
We first scale each bin capacity proportionally so that the total fits within $M$ tiles, i.e., $|B_s| \leftarrow \lfloor |B_s| \cdot M / \sum_{s'} |B_{s'}| \rceil$ (Figure~\ref{fig::guided_solver2}b), and then repack tasks within each bin.
\revision[blue][R2-Q4]{%
As shown in Figure~\ref{fig::guided_solver2}b, the combined bin (\textbf{2\&4}) is repacked, where task B2 is reshaped to fit the reduced capacity, yielding a more compact layout.
}
We adopt a first-fit-decreasing style heuristic with iterative improvement: (1) sort items by a tie-broken priority (deadline/criticality, then index); (2) place each item at the earliest feasible offset that respects precedence and bin capacity; (3) iterate to compact gaps and reduce peak usage. The procedure returns, for every bin, the final size and position of each item.

\subsubsection{Physical Partition Binding}\label{sec::physical_partition}
\revision[blue][R1-Q4d,R4-Q5]{Phase III yields the final logical bin capacities $|B_s|$ and item-to-bin
mappings. To map these logical bins to physical tiles in rectangular areas,
we adopt the classical Guillotine cutting algorithm~\cite{beasley1985algorithms},
a heuristic that divides a large rectangle into smaller ones of target sizes through a series of bisecting end-to-end cuts.
This process also maps each partition to its nearest memory controller (cf.\ Section~\ref{sec::hw_model}).
This mapping minimizes cross-partition NoC traffic and establishes fixed data paths for normal execution.}

%% case_study.tex
%% Section IV-C: Case Study --- Limitations of Existing Schedulers
%% Section IV-D: Opportunity and Challenges
%% Extracted from 0-motivation-zyp.tex L264-340, per sec2-4_COT_v3.md

\subsection{Case Study: Limitations of Existing Schedulers}\label{subsec:limitations}

\begin{figure*}[t]
  \centering  
  \begin{minipage}[c]{0.32\linewidth}
    \centering
    \subfloat{
      \mytikzmark{a} \includegraphics[width=\linewidth]{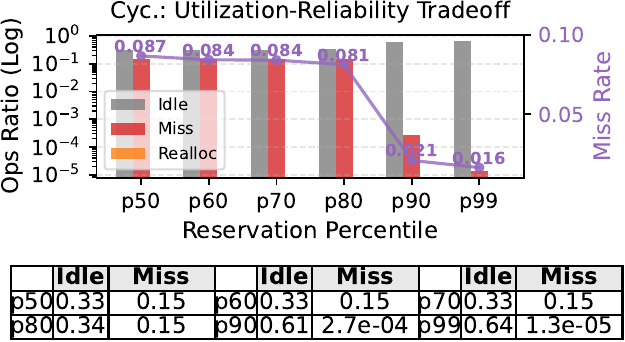}
      \label{fig::case1_tradeoff}
    }
  \end{minipage}
  \begin{minipage}[c]{0.32\linewidth}
    \centering
    \subfloat{
      \mytikzmark{b} \includegraphics[width=\linewidth]{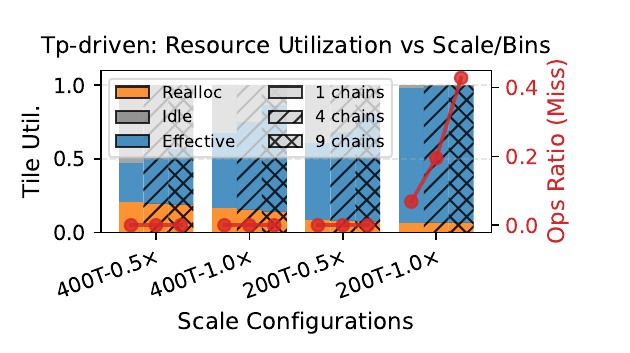}%
      \label{fig::case2_utilization}
    }
    \end{minipage}
    \begin{minipage}[c]{0.32\linewidth}
      \centering
      \subfloat{
        \mytikzmark{c} \includegraphics[width=\linewidth]{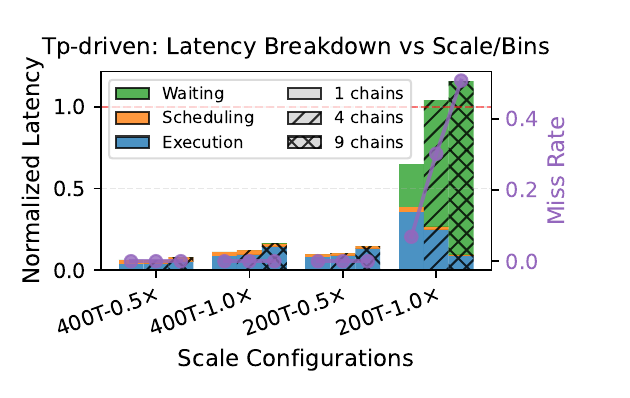}%
        \label{fig::case2_breakdown}
      }
      \end{minipage}
    \begin{tikzpicture}[overlay,remember picture]
    \node [xshift=-0.2cm,yshift=0.3cm] at (a) {(a)}; %[xshift=1cm,yshift=0cm] 
    \node [xshift=0.5cm,yshift=0.25cm] at (b) {(b)}; %[xshift=1cm,yshift=0cm] 
    \node [xshift=0.3cm,yshift=0.4cm] at (c) {(c)}; %[xshift=1cm,yshift=0cm] 
  \end{tikzpicture}
  \caption{
    \small{Characterization of \CyclicName and \TpFirstName on a \hwname ADS.
    \revisioninline[blue][R2-Q5]{%
    Total \textbf{processing power} is normalized to 1, where total capacity is decomposed into \highlightuline{idle}, \highlightuline{effective} use, and waste due to reallocation (\highlightuline{Realloc.}). For cases in which some tasks time out, \concept{Miss Rate} measures the percentage of timed-out tasks, and \concept{missed processing power ratio} measures the ratio of dropped processing power to total processing power.%
    }%
    \textbf{(a)}~\CyclicName: left axis decomposes capacity into \highlightuline{idle}, \highlightuline{miss},
    and realloc fractions (log scale) swept over percentile~$q$;
    right axis shows per-task miss rate.
    \textbf{(b)}~\TpFirstName: resource utilization breakdown,
    clustered by hardware scale (i.e., tile count) and workload scale ($\times$ chain count, load factor);
    the purple line shows miss rate.
    \textbf{(c)}~\TpFirstName: end-to-end latency breakdown normalized to $\mathcal{D}_{e2e}$.}
  }\label{fig::movit-exp-result}
  \vspace{-1em}
\end{figure*}
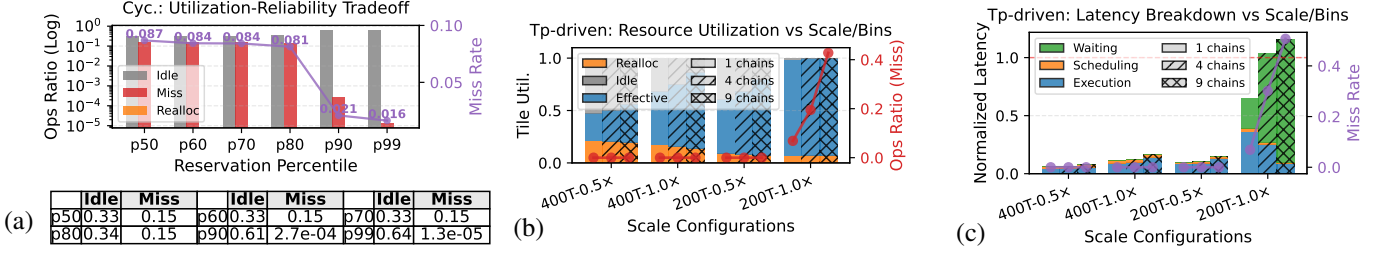

With GHA producing baseline schedules for both paradigms, we conduct a case study to reveal their fundamental limitations under runtime variation. The experiments are implemented with a fast event-driven simulator, Tile-stream (the full experimental setup appears in Section~\ref{sec::exp_setup}), evaluating both schedulers under
execution and workload variations.
% Using GHA (Section~\ref{sec::GHA}) to produce scheduling plans, per-task dataflow mapping by CoSA~\cite{huang2021cosa},

\subsubsection{\CyclicName}

\revision[blue][R4-Q1]{%
We sweep the quantile $q$ to study how per-task provisioning affects utilization and reliability.
Figure~\ref{fig::movit-exp-result}a decomposes total processing power (left axis) into idle, missed, and reallocation fractions, all normalized to total capacity per hyper-period; the right axis shows the per-task miss rate.
Two observations stand out.
First, raising $q$ reduces the miss rate but inflates the idle fraction---\CyclicName cannot achieve high utilization and low miss rate simultaneously.
Second, for $q \geq 0.9$ the idle fraction far exceeds the fraction of dropped workload, revealing that the fixed-DoP strategy pays a disproportionate resource cost for marginal reliability gains.}

\textbf{Analysis.}
Both observations stem from the same root cause: per-task budget isolation is structurally misaligned with the E2E nature of ADS.
In our profiled workloads, provisioning every task with its $q$-quantile budget yields a measured chain-level timeout rate far below $(1{-}q)$.
This indicates that task-level tail events rarely align within the same chain instance; therefore, summing individual tail budgets gives a conservative estimate of the E2E tail.
This empirical conservatism corresponds to the tail-composition headroom discussed in Section~\ref{sec::latency_model}. 
Yet \CyclicName's strict isolation forecloses cross-task slack sharing and cannot reclaim this headroom: as shown in Figure~\ref{fig::sched_space_plus_tab}e, \singlefunc-A exceeds its budget under F1 (Section~\ref{sec::workload_model}) and is terminated, while adjacent tasks sit idle with guaranteed but unused time windows.
Overcoming this pessimistic per-task provisioning cost is a key motivation for \PaperName's runtime sharing mechanism (Section~\ref{sec::hybrid-grained_isolation}).

\subsubsection{\TpFirstName}

\revision[blue][R4-Q1]{%
We sweep hardware scale and workload scale to study the scalability and inter-task interference
of \TpFirstName, collecting utilization breakdown with respect to total processing power
(Figure~\ref{fig::movit-exp-result}b) and end-to-end latency breakdown normalized to
$\mathcal{D}_{e2e}$ (Figure~\ref{fig::movit-exp-result}c).
Hardware scale varies over tile count $\{200, 400\}$; workload scale varies over number of replicated \aux chains
$\{1, 4, 9\}$ and load factor $\{0.5, 1\}$.
Although the rescheduling delay is small in a single \ETEs chain, the processing power wasted
across all tiles is significant (10\%--20\% idle at small and medium scale).
Two trends emerge: when utilization is high the rescheduling overhead is high; as load increases
or hardware shrinks, rescheduling overhead falls while queuing delay and miss rate rise.}
Under the same load, a larger hardware scale produces higher rescheduling overhead.

\textbf{Analysis.}
\revision[blue][R3-Q2]{The root cause is the absence of an E2E view: work-conserving allocation always exhausts
all tiles, so every arriving high-priority task triggers a full rescheduling.}
In ADS, with dozens of DNNs running at 10--240\,Hz, each task instance creates two scheduling events,
on arrival and completion, so the runtime can face thousands of reallocation opportunities per second.
In contrast, \TpFirstName schedulers are designed to handle low request 
rates (typically $\leq$10 requests per second (RPS) per chip~\cite{Jiang2018a-Mainstream}), and                                                                    
batch processing amortizes the cost. 
A full-chip tile migration on a high-performance accelerator
($\times$10\,MB SRAM, $\times$100\,GB/s bandwidth~\cite{Tenstorrent-2022-ISSCC}) costs hundreds of
microseconds; these delays compound along multi-stage E2E chains.
As shown in Figure~\ref{fig::sched_space_plus_tab}f, even with full tile utilization,
repeated rescheduling stalls cause \singlefunc-B's
E2E timeout.

\subsection{Opportunity and Challenges}\label{subsec:opportunity}

The above observations call for a bounded spatio-temporal scheduling space that
can reclaim DAG slack and exploit DoP tunability without repeated global
reallocation.
First, the scheduler must support colocation, adaptive sharing, and interference
isolation. 
Second, an effective scheduler must be aware of the costs and benefits of rescheduling and
isolation, triggering reallocation only when the latency benefit
outweighs the cost.
These observations motivate \PaperName, which retains GHA as the offline compiler and
introduces a runtime scheduler with controlled elasticity, detailed in
Section~\ref{sec::ADS-Tile}.

% Section V: ADS-Tile Runtime Scheduler
\section{\PaperName Runtime Scheduler} \label{sec::ADS-Tile}

This section presents the \PaperName DAG-aware runtime scheduler. We first give a framework overview showing how offline and runtime components interact, then detail spatio-temporal isolation-sharing space and how the DAG-aware runtime scheduler uses it.
% that bound runtime reallocation

\subsection{Overview}
\begin{figure}[ht]
    \centering
    \begin{minipage}[b]{\linewidth}
      \centering
      \subfloat{
        \includegraphics[width=0.8\linewidth]{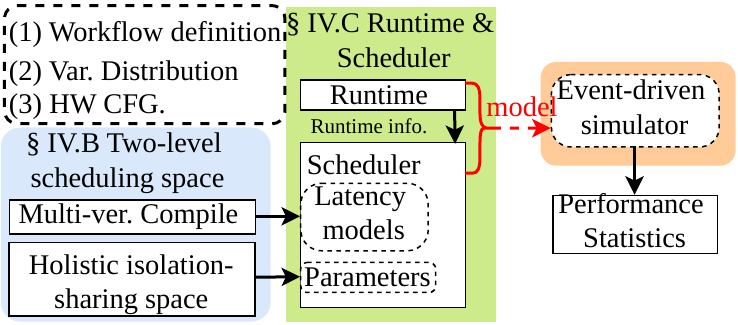}
      }
    %   \hfill
    %   \subfloat{
    %     \includegraphics[width=0.2\linewidth]{figures/pdf/sched_arch.drawio.pdf}
    %   }
      \caption{Overview of the \PaperName framework.}%
      \label{fig::overview}
    \end{minipage}
    \vspace{-1.5em}
\end{figure}

As shown in Figure~\ref{fig::overview}, \PaperName takes a workflow graph, hardware configuration, and variation distributions as input, and operates in two stages.
\emph{Offline}, the GHA compiler (Section~\ref{sec::GHA}) produces a static schedule---partition assignment $x_{vs}$, reservation parameters $(t_v, ddl_{\text{sub},v})$ of each task and layout of partitions, which serves as the baseline operating point. % per-task spatial sizes $c_v$, latency budget $l_v$,
\emph{At runtime}, a distributed scheduler running on each partition's RISC-V controller (Section~\ref{sec::reservation_runtime}) dynamically adjusts tile allocations around this baseline.
The scheduler operates within two coordinated mechanisms: configurable isolation bounds \emph{where} reallocation propagates (\cref{sec::hybrid-grained_isolation}), while elastic reservation controls \emph{when} tasks become eligible for colocation.
These constraints are detailed next.

\subsection{Spatio-temporal Isolation-sharing Space}~\label{sec::hybrid-grained_isolation}
To exploit DoP tunability in colocation scheduling without incurring high
scheduling-induced overhead, \PaperName combines the strengths of the two
baseline paradigms.
As observed in \cref{sec::non-isolated_spatial-aware_scheduler}, directly
adopting the work-conserving policy of \AbbrCallout{\TpFirstName} can lead to
chip-wide and frequent reallocation.
Our key observation is that the spatio-temporal isolation used by
\AbbrCallout{\CyclicName}-like schedulers can be decomposed into two separable
constraint dimensions:
\begin{enumerate*}[label=(\roman*)]
    \item \emph{Isolation} constrains the spatial scope, specifying
    \emph{where} tiles may be redistributed.
    \item \emph{Reservation} constrains the temporal scope, specifying
    \emph{when} tasks may be admitted for scheduling.
\end{enumerate*}
We combine each dimension with \TpFirstName-style colocation-aware dynamic
scheduling through the two mechanisms below.
Together, the two constraints form a spatio-temporal isolation-sharing space:
adaptive tile allocation remains available, but reallocation is bounded in space
and time.
\begin{figure}[htbp]
  \centering
  \includegraphics[width=0.8\linewidth]{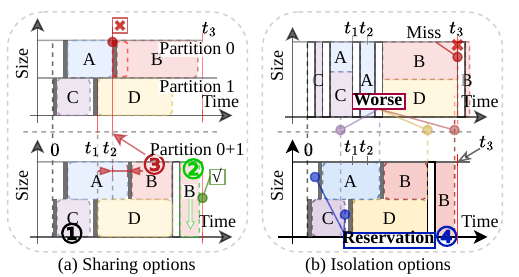}
  \caption{Spatio-temporal isolation-sharing space for extending (a) \CyclicName and (b) \TpFirstName:
  \ding{172}: coarsen partitions to allow colocation, \ding{173}: allocate tiles adaptively among co-located tasks, \ding{174}: use soft sub-deadlines for intermediate tasks and share E2E slack released along DAG edges,
  \ding{175}: reserve idle tiles for incoming tasks.
  \revisioninline[blue][R2-Q6]{In~(b), task-A arrives late due to upstream jitter; both A and B experience load burst.}
  }
  \label{fig::reschedule_policies_counterparts}
  \vspace{-0.5em}
\end{figure}

\subsubsection{Configurable isolation} 
The first insight is to avoid chip-wide reallocation by using the GHA-generated
partitions as spatial scheduling boundaries
(\cref{subsec::ads_platform_evolution}).
Instead of enforcing per-chain or per-task isolation,
\ding{172} tasks from different \ETEs chains are allowed to be colocated into the same partition, as shown in Figure~\ref{fig::reschedule_policies_counterparts}a.
The colocation-aware dynamic scheduler can then
allocate tiles adaptively among co-located tasks, but only within the partition-local tile pool.
Given the partition layout and task assignment (from GHA), each partition forms an
independent colocation domain with its own task queue and tile pool.
No task can migrate across partition boundaries, and idle tiles in other
partitions cannot be used by the current partition.
Consequently, each rescheduling event only affects tasks in its partition,
bounding the data-migration cost of a single reallocation.
By controlling the partition granularity, the scheduler balances isolation and sharing in space.

\subsubsection{Elastic reservation}
The second insight is to reduce unnecessary reallocation by reserving idle tiles for urgent arrivals when overall load is low, avoiding reallocation when those tasks arrive.
% This mechanism avoids a fully work-conserving policy in time: tasks before their ERT are not eligible for colocation, and not all idle tiles are immediately assigned to currently active tasks.
We implement this via two controls:
\begin{enumerate*}[label=(\roman*)]
    \item (Admission Control) Each task is assigned a predefined Earliest-Ready-Time (ERT, $t_v$). A task is not eligible for colocation until its ERT is reached.
    \item (Quota Control) Each task's sub-deadline $ddl_{\text{sub},v}$ serves as the target finish time for quota estimation. The allocation policy selects the minimum tile quota $c_v^{(t)}$ expected to finish the task before this target,
\end{enumerate*}
leaving unused tiles idle rather than ``distributing all spare tiles'' as in \AbbrCallout{\TpFirstName}.
These two parameters are also generated by GHA.

The reservation parameters $(t_v, ddl_{\text{sub},v})$ are derived by solving
the same formulation (Eq.~\ref{eq:phase1_obj}--\ref{eq:phase1_core}).
A smaller $q$ yields shorter per-task latency budgets, advancing both $t_v$ and
$ddl_{\text{sub},v}$ and thus tightening the reservation window.
As shown in Figure~\ref{fig::reschedule_policies_counterparts}b (bottom~\ding{175}),
when allocating tiles for tasks C and A, the scheduler also reserves idle tiles
for incoming tasks A and D.
This eliminates four rescheduling events and improves overall latency;
only the rescheduling for task B is retained because its latency gain outweighs the migration cost.
This trades some utilization in the current scheduling window for lower future timeout risk.

\subsection{DAG-aware Runtime Scheduler} \label{sec::reservation_runtime}
% \PaperName runtime is response to deploy scheduling algorithms on distributed tiles, and provides support to scheduler in determining the best choices at each scheduling point.
% The runtime implements the proposed scheduling algorithm as a two-level scheduler (Figure~\ref{fig::rt_glb}), consisting of a host part and multiple local parts that work in parallel, executed on host processor and the RISC-V controller first tile of each partition, respectively.

% The Host scheduler maintains partition.
% Given a reservation table, the host scheduler first initializes partitions based on the partition scheme,
% as well as the routing table of streaming data among sensor processing units, main memory and partitions based on the mapping scheme and subscription relationship.
% Local part consists of a progress monitor (PM), local scheduler.
% In each partition (Figure~\ref{fig::rt_loc}), a PM tracks the task's state and progress.
% % Based on the received data and predefined matching logic, for each task, the PM can track its event time, absolute deadline of its belonged pipeline ($Getddl$), and determine whether it is activated ($CheckAct$).
% The received data are cached on stream buffer, sorted by their event time and their source task.

\begin{figure}[htbp]
    \centering  
    \begin{minipage}[c]{0.45\linewidth}
      \centering
      \subfloat{
        \mytikzmark{a} \includegraphics[width=\linewidth]{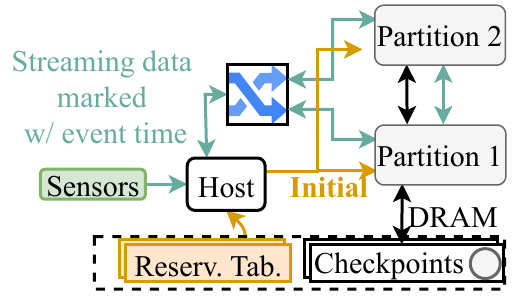}
        \label{fig::rt_glb}
      }
    \end{minipage}
    \hfill
    \begin{minipage}[c]{0.45\linewidth}
      \centering
      \subfloat{
        \mytikzmark{b} \includegraphics[width=\linewidth]{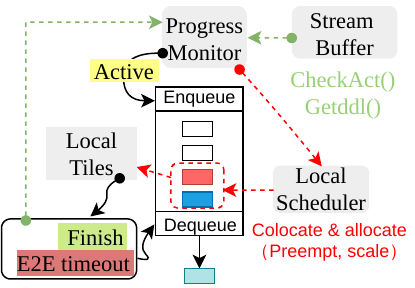}%
        \label{fig::rt_loc}
      }
      \end{minipage}
    \begin{tikzpicture}[overlay,remember picture]
      \node [xshift=0.5cm,yshift=-0.2cm] at (a) {(a)}; %[xshift=1cm,yshift=0cm] 
      \node [xshift=-0.5cm,yshift=0.05cm] at (b) {(b)}; %[xshift=1cm,yshift=0cm] 
    \end{tikzpicture}
    \caption{Flowchart of the runtime scheduler.
    }%
    \vspace{-1em}
  \end{figure}
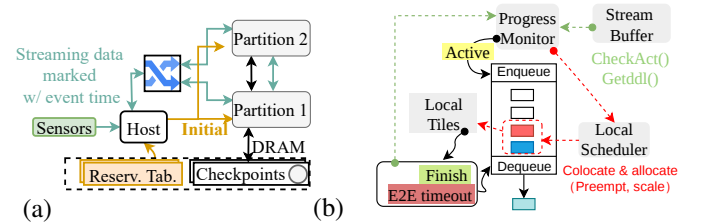
The \PaperName runtime executes the proposed scheduling algorithm across
distributed tiles through a two-level scheduling architecture
(Figure~\ref{fig::rt_glb}).
A host scheduler initializes partitions using the reservation table and
configures data routing between sensor processing units, main memory, and
partitions.
Each partition then runs a local scheduler on the RISC-V core of its first tile.
The local scheduler consists of a progress monitor (PM) and a task coordinator.

Within each partition (Figure~\ref{fig::rt_loc}), incoming data are cached in a
stream buffer and sorted by event time and source task index.
The PM tracks each task's state and progress.
Tasks are activated when all predecessor dependencies are satisfied
($CheckAct$) and enqueued to $Q_{\text{active}}$.
For tasks receiving more than one stream, the PM aligns inputs using the
predefined event-time matching logic~\cite{Gog2022-Eurosys-D3,
Luo2019-UCB_safeos}.
From the event time and relative timing constraints, the PM derives the
absolute deadlines of individual \singlefuncs and their \ETEs chains
($Getddl$).
Tasks are dequeued from $Q_{\text{active}}$ when completed or when they miss the
\ETEs deadline ($ddl_{\text{e2e}}$).

Within this partition-local boundary, the scheduler exploits the DAG structure
through two forms of sharing.
First, it enables spatial sharing across co-active paths: admitted tasks in the
same partition share the tile pool, and tiles are allocated according to urgency
(\ding{173} in Figure~\ref{fig::reschedule_policies_counterparts}a).
Second, it enables temporal sharing along DAG edges: sub-deadlines are treated
as soft references rather than hard stop points, so a delayed task may consume
slack from adjacent stages when the E2E deadline still permits
(\ding{174} in Figure~\ref{fig::reschedule_policies_counterparts}a).
Together, cross-path tile sharing and within-path slack sharing reduce the
conservatism of per-task quantile bounds
(Section~\ref{sec::latency_model}) while keeping reallocation confined to the
current partition.

\revision[blue][R3-Q6]{Algorithm~\ref{alg:reschedule} implements this
DAG-aware sharing policy.}
The local scheduler decides which tasks can colocate in a partition and how
tiles are allocated among them, using the offline priors in the scheduling
table, the constraints imposed by configurable isolation and elastic
reservation, and the task states maintained by the PM.
The scheduling table provides the two reservation parameters generated offline:
the ERT and sub-deadline of each task.
At each scheduling point, each local PM synchronizes task completion states,
refreshes the active queue through $\textsf{CheckAct}$, and admits
only tasks whose ERT has arrived (line~\ref{alg:runtime:admission}) into the ready list ($Q_{\text{ready}}$).
The $\textsf{ChkTrigger}$ (line~\ref{alg:runtime:trigger}) uses newly activated tasks, runtime free tiles, and the
current time to decide whether to reschedule the running tasks.
When the trigger fires, quota control orders admitted tasks by sub-deadline and uses $\textsf{FitQuota}$
to select the smallest feasible DoP candidate based on each task's remaining load and deadline slack
(lines~\ref{alg:runtime:sort}--\ref{alg:runtime:fitquota}).
The selected DoP targets completion by the task's sub-deadline when sufficient free tiles are available,
without overusing the available tiles. 
Residual $\textsf{Cap}$ is left idle for future urgent arrivals before PM applies the new
partition-local allocation (line~\ref{alg:runtime:reallocate}).
\begin{algorithm}[t]
  \caption{(Partition) DAG-aware colocation and allocation}
  \label{alg:reschedule}
  \footnotesize
  \begin{algorithmic}[1]
    \Require Partition map $\mathsf{PartMap}$; scheduling table
    $\mathsf{SchedTab}[v]=(\mathrm{ERT}:t_v,\mathrm{DDL}:ddl_{\text{sub},v})$;
    current time $t_{\mathrm{curr}}$
    \Ensure Updated partition-local allocation map $\mathsf{AllocMap}$

    \State $new\_active, Q_{\mathrm{active}} \gets$ PM.CheckAct()
    \label{alg:runtime:checkact}

    \LComment{\color{blue}-----\concept{Admission Control}: admit by ERT-----}
    \State $Q_{\mathrm{ready}} \gets
      \{v \in Q_{\mathrm{active}}
      \mid \mathsf{SchedTab}[v].\mathrm{ERT} \leq t_{\mathrm{curr}}\}$
    \label{alg:runtime:admission}

    \If{\textbf{not} ChkTrigger($new\_active$, PM.GetFreeTiles(), $t_{\mathrm{curr}}$)}
      \State \Return PM.GetAllocMap()
    \EndIf
    \label{alg:runtime:trigger}

    \LComment{\color{blue}--------\textbf{Quota Control}: DDL order with reserved residual capacity------}
    \State $Q_{\mathrm{ready}} \gets
      \operatorname{sort}^{\uparrow}_{\mathsf{SchedTab}[\cdot].\mathrm{DDL}}
      (Q_{\mathrm{ready}})$
    \label{alg:runtime:sort}
    \State $\mathsf{AllocMap} \gets \emptyset$,
           $Cap \gets$ PM.GetCapacity()
    \label{alg:runtime:initcap}

    \For{$v \in Q_{\mathrm{ready}}$ \textbf{while} $Cap > 0$}
    \label{alg:runtime:stop}
      \State $ld_v \gets$ PM.GetLoad($v$),
             $\mathcal{D}_v \gets$ PM.GetSlack($v,t_{\mathrm{curr}}$)
      \label{alg:runtime:params}

      \State $c_v^{(t)} \gets$
             FitQuota($v$, $ld_v$, $\mathcal{D}_v$, $Cap$)
      \label{alg:runtime:fitquota}

      \If{$c_v^{(t)} > 0$}
        \State $\mathsf{AllocMap}[v] \gets c_v^{(t)}$, $Cap \gets Cap - c_v^{(t)}$
      \EndIf
    \EndFor

    \Statex \Comment{Residual $Cap$ is reserved for incoming tasks.}

    \State PM.Reallocate($\mathsf{AllocMap}$)
    \label{alg:runtime:reallocate}
    % \State PM.UpdateQueue($\mathsf{AllocMap}$, $\mathsf{PartMap}$)
    % \label{alg:runtime:updatequeue}
    \State \Return $\mathsf{AllocMap}$
    \vspace{-2em}
  \end{algorithmic}
\end{algorithm}

\subsection{Implementation Details}
\label{sec::impl_details}
The above algorithm determines which tasks to co-locate and how many
tiles to allocate. We now detail the implementation that makes these decisions practical with minimal overhead.

\subsubsection{Switching Precompiled Implementations}
\label{sec::impl_switching}
First, all tasks in a partition are stalled, regardless of whether they need to be rescheduled.
For the tasks that need to be rescheduled, their partial results in the SRAM of the original tiles are checkpointed, 
then the results are re-sharded and migrated to new tiles according to the pattern determined at compile time,
and finally the execution of all tasks in the partition is resumed.

\subsubsection{Multi-version Compilation and Dataflow Mapping}
\label{sec::multi_version_impl}
Each DNN model is compiled into an executable program. During compilation, we search for the optimal dataflow mapping, including loop tiling, loop ordering, and loop unrolling. 
Each DoP candidate is profiled under varying workload intensities to construct the chunk-level probabilistic latency model $L_v(q, c_v)$, used by both the GHA solver (Section~\ref{sec::GHA}) and the runtime scheduler. 
To avoid retaining too many DoP candidates, we gradually increase the tile count from the minimum and prune candidates that do not improve latency by at least a threshold over the previous candidate.
Compilation proceeds at operator-chunk granularity.
Each chunk spans a contiguous operator segment, allowing a DNN task to switch its DoP during execution. We restrict switching to chunk boundaries, which lets the compiler pre-compute all resharding traffic offline: for every pair of DoP candidates across adjacent chunks, it statically maps the output layout of the preceding chunk to the input layout of the current chunk. 
Each chunk runs as an unpreemptable unit; switching therefore requires neither register backup nor runtime compilation of migration traffic.

% ==== Chinese draft preserved for reference ====
% 首先，每一个模型会被编译为一个程序。在这期间，会搜索最优的是数据流，包括循环的切分、循环的排序以及循环的展开。
% 每个模型会在各种负载条件下构建概率的延迟模型，这一模型会被GHA求解器或者运行时调度器使用。
% 为了避免生成太多的版本，我们会从最少的参数数开始增加资源，并剪枝掉那些延迟提升不明显的版本。
% 值得注意的是，编译的流程以 chunk 为力度进行，每个 chunk 是 DNN 模型的一部分算子。
% 这么做是为了支持模型在部分执行的时候可以切换并行度。为了避免运行时，计算复杂的traffic。切换仅被允许发生在 chunk 边界的地方，每个 chunk 都是一个不可抢占的执行单元。
% 每个 chunk 而言，都包含从前一个chunk的各个DoP 版本的输出layout，到当前chunk的各个DoP 版本的输入layout的resharding 的 traffic 数据流。

\subsubsection{Dynamic Tile Mapping within Partitions}
\label{sec::dynamic_mapping}

The mapping from logical tiles to physical tiles is divided into two stages.
First, the physical tiles belonging to each partition are determined statically
at compile time using Guillotine cutting (Section~\ref{sec::physical_partition}).
Second, the mapping within each partition is determined dynamically at runtime.
% We adopt a virtual mapping layer inspired by AuRORA~\cite{kim2023aurora}.
% This indirection allows the runtime to place a task's logical tiles onto any
% available physical tiles within the partition, minimizing data movement during
% rescheduling. When a task's tile allocation changes, the virtual mapping layer
% computes a new physical placement that minimizes the distance from the previous
% configuration, reducing migration overhead.
We introduce a logical-to-physical (L2P) tile mapping mechanism, inspired by AuRORA~\cite{kim2023aurora}. 
This indirection decouples a task's logical representation from its physical execution, allowing the runtime to flexibly map logical tiles to any available physical tiles within the partition. 
During rescheduling, the runtime computes a new placement that minimizes the distance from the previous configuration and updates the translation table, effectively reducing data migration overhead.

\section{Evaluation}% 
\label{sec::eval}

\subsection{Experimental Setup}%
\label{sec::exp_setup}

\textbf{Hardware Configuration.}
\revision[blue][R4-Q1]{%
We target a \hwname architecture based on Simba~\cite{Nv_simba}.
Each chip contains 128 tiles operating at 2\,GHz, connected by a 2D mesh NoC.
The on-chip SRAM is distributed across tiles, with 1.25\,MB per tile. Each inter-tile NoC link is 64\,B wide. 
Each tile consists of 16 PEs. 
Each PE contains 16 16-bit multiplier-accumulator units organized in weight-stationary (NVDLA) dataflow~\cite{NVIDIA2018}.
Off-chip memory is 128-bit LPDDR5 DRAM with 102\,GB/s bandwidth. 
Multiple chips are connected via PCIe Gen5 x4 links (15.8\,GB/s/direction).}

\begin{figure*}[t]
    \centering
    \hspace*{-3.5cm}
    \begin{minipage}[]{0.8\linewidth}
          \begin{Overpic}[]{%

\resizebox{\linewidth}{!}{
  \begin{tabular}{lll>{\color{blue}}r>{\color{blue}}r|lll>{\color{blue}}r>{\color{blue}}r}
    \toprule
    ID & Task & Model & {\begin{tabular}[c]{@{}r@{}}Avg. \\BW(\%)\end{tabular}} & {\begin{tabular}[c]{@{}r@{}}Peak\\(GB/s)\end{tabular}} &
    ID & Task & Model & {\begin{tabular}[c]{@{}r@{}}Avg. \\BW(\%)\end{tabular}} & {\begin{tabular}[c]{@{}r@{}}Peak\\(GB/s)\end{tabular}} \\
    \midrule
    1 & \begin{tabular}[c]{@{}l@{}}Traffic light \\detection\end{tabular} & \begin{tabular}[c]{@{}l@{}}Resnet18 (E)~\cite{resnet15}+\\brake classifier~\cite{Chen2022-LAV} \end{tabular} & 8.4 & 14.4 &
    7 & \begin{tabular}[c]{@{}l@{}}Steering \\\& Speed control \end{tabular} & LAV~\cite{Chen2022-LAV} & 0.1 & 2.0 \\
    \cline{1-10}

    2 & Image backbones & YoloX (E)~\cite{Ge2021-YOLOX} & 50.7 & 17.1 &
    8 & \begin{tabular}[c]{@{}l@{}}Stereo-LiDAR \\fusion\end{tabular} & \begin{tabular}[t]{@{}l@{}}ERFNet (E)+ Point-\\Painting~\cite{Bergasa2018_ERFNet, Vora2020-Pointpainting}\end{tabular} & 5.4 & 21.0 \\
    \cline{1-10}

    3 & \begin{tabular}[c]{@{}l@{}}Multi-camera \\fusion\end{tabular} & BevFormer (E)~\cite{Li2022-BEVFormer} & 19.0 & 280.2 &
    9,11,12 & \begin{tabular}[c]{@{}l@{}}Lane/drivable area/\\Semantic \\segmentation\end{tabular} & ERFNet (H)~\cite{Chen2022-LAV, Bergasa2018_ERFNet} & 2.5--4.9 & 26.8--27.2 \\
    \cline{1-10}

    4 & \begin{tabular}[c]{@{}l@{}}Visual object \\detection\end{tabular} & \begin{tabular}[c]{@{}l@{}}Deformable \\DETR (H)~\cite{Zhu2020-Deformable_DETR}\end{tabular} & 1.7 & 31.9 &
    10 & \begin{tabular}[c]{@{}l@{}}LiDAR-based \\detection\end{tabular} & \begin{tabular}[t]{@{}l@{}}PointPillars~\cite{Lang2019_Pointpillars}\\CenterNet (H)~\cite{Yin2021-centerpoint}\end{tabular} & 1.2 & 78.2 \\
    \cline{1-10}

    5 & \begin{tabular}[c]{@{}l@{}}Trajectory \\prediction\end{tabular} & \multirow{2}{*}{LAV~\cite{Chen2022-LAV}} & 1.3 & 10.3 &
    13 & Optical Flow & PWC-NET (H)~\cite{Sun2018-PWC-Net} & 1.0 & 4.8 \\
    \cline{1-2}\cline{4-10}

    6 & Path planning & & 1.3 & 1.0 &
    14 & Depth estimation & SemAttNet (H)~\cite{nazir2022semattnet} & 2.5 & 15.3 \\
    \bottomrule
    \multicolumn{10}{l}{\small $\bullet$ H: head, E: encoder/backbone. \revisioninline[blue][R3-Q3]{Avg.\ BW = average fraction of aggregated LPDDR5 bandwidth; Peak = instantaneous maximum.}}
    \end{tabular}
  }

              }%
              \put(100,2){\includegraphics[width=0.2\linewidth]{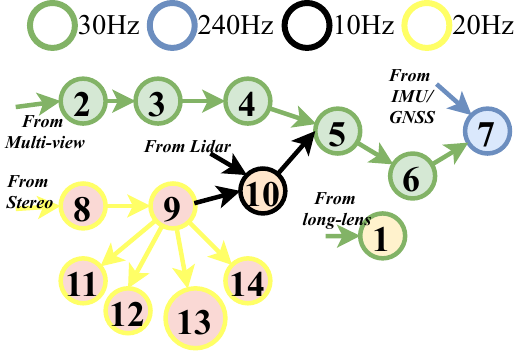}}
          \end{Overpic}
    \end{minipage}
  \caption{
     The graph abstraction and algorithms adopted for the L4 ADS benchmark.
  }%
  \label{fig::e2e_workflow_abs_plus_table}
  \vspace*{-0.5cm}
\end{figure*}

\textbf{Benchmarks.}
\revision[blue][R4-Q1]{%
We adopt an L4 ADS benchmark derived from industry and academia workloads~\cite{resnet15, Ge2021-YOLOX, Li2022-BEVFormer, Zhu2020-Deformable_DETR, Chen2022-LAV, Bergasa2018_ERFNet, Lang2019_Pointpillars, Sun2018-PWC-Net}.
The sensing system includes multi-view cameras (30\,Hz), stereo cameras (20\,Hz), LiDAR (10\,Hz), and IMU (240\,Hz).
Figure~\ref{fig::e2e_workflow_abs_plus_table} shows the DAG topology and per-task model implementations.
To evaluate scalability, we scale the number of concurrent \ETEs chains by replicating \aux pipelines (nodes 11--14 in Figure~\ref{fig::e2e_workflow_abs_plus_table}).
The \ETEs latency constraint is set to 80\,ms, 90\,ms, and 100\,ms for safety-critical paths across different scenarios, and 100\,ms for non-critical paths.}
% To generate cases of different dynamics, our benchmark supports generating variable load by switching DNN models with different number of region proposed for object detection, predicted trajectories for prediction, and planned paths for planning. 

\revision[blue][R3-Q3]{%
The benchmarks require roughly 180--300\,TMAC/s of FP16 compute, corresponding to 3--5 accelerator chips under our hardware configuration.
Figure~\ref{fig::e2e_workflow_abs_plus_table} reports the per-task memory bandwidth profile, where image backbones and multi-camera fusion dominate because they process multi-camera streams.
Under our multi-chip configuration, these bandwidth-intensive models can be allocated to tiles across multiple chips via pipeline or data parallelism, so that the aggregate bandwidth meets the requirement.
Bandwidth contention arises when multiple high-bandwidth tasks access DRAM simultaneously, which is captured by our latency model (Section~\ref{sec::latency_model}).}

\textbf{Baselines.}
We compare \PaperName against two representative scheduling paradigms:
\begin{enumerate*}[label=(\arabic*)]
    \item \textbf{\CyclicName} (fully-isolated time-multiplexing): Each task receives a fixed tile count and time slot derived from WCET analysis~\cite{baker1989cyclic}.
    \item \textbf{\TpFirstName} (work-conserving dynamic scheduling): Tiles are greedily reallocated among active tasks to maximize utilization, following Planaria~\cite{Ghodrati2020-Planaria}.
\end{enumerate*}

\textbf{Simulation Methodology.}
\revision[blue][R4-Q1]{Our evaluation employs a two-level simulation hierarchy.
\emph{(1) Architecture-level performance model.}
Per-operator latency tables are produced offline.
We use CoSA~\cite{huang2021cosa} to generate dataflow mappings for each DNN operator under varying tile counts, built on top of Timeloop~\cite{Timeloop}, which provides cycle-accurate latency models calibrated against hardware measurements.
For NoC and DRAM bandwidth contention, we inject packet-level traffic following the ADS workflow patterns into a 2D-mesh NoC and simulate it with BookSim~\cite{dally2004principles}. 
We sweep the network load from almost idle to saturation, and fit the resulting per-hop latency distributions to obtain the stochastic $I_v$ parameters (Section~\ref{sec::latency_model}).
\emph{(2) Event-driven system simulator.}
Tile-stream is an event-driven simulator that models streaming data from sensors, packet movement among tiles, and scheduler decisions at microsecond granularity.
It takes the DAG topology, sensor configurations, and pre-characterized per-operator latency tables as input, and reports per-task progress, resource occupancy, and \ETEs latency distributions under realistic arrival patterns and variation factors.
%
% \emph{(3) Overhead modeling.}
The \emph{rescheduling overhead} in Tile-stream consists of three components: (1)~scheduler decision ($<$10\,$\mu$s on the RISC-V controller, negligible), (2)~context switch (state checkpoint to DRAM), and (3)~data migration via NoC (dominant, proportional to checkpoint size).
Migration latency is derived from the 2D-mesh NoC hop latency and LPDDR5 bandwidth, scaled linearly by the checkpoint data volume~\cite{Nv_simba, Kim2023MoCA}.}

\subsection{Ablation Study}
\label{sec::ablation}

\begin{figure*}[ht]
  \centering
  \begin{minipage}[t]{0.25\linewidth}
    \centering
    \includegraphics[width=\linewidth]{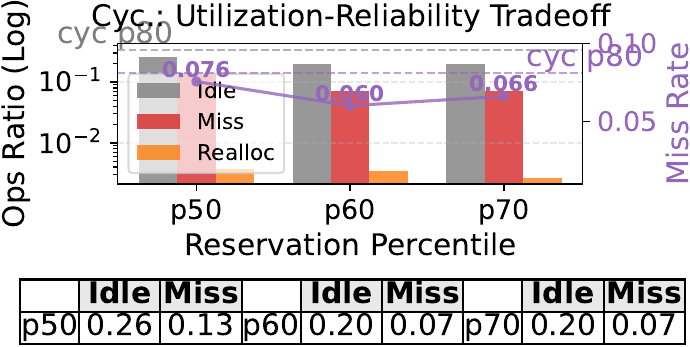}\\
    \small (a) Dynamic reservation (cyc-S vs.\ cyc)
  \end{minipage}
  \hfill
  \begin{minipage}[t]{0.24\linewidth}
    \centering
    \includegraphics[width=\linewidth]{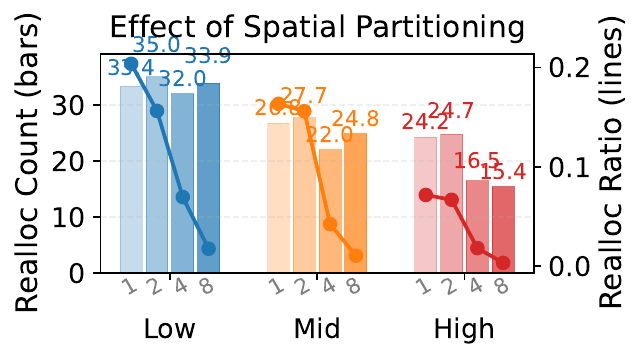}\\
    \small (b) Partitioning: rescheduling overhead
  \end{minipage}
  \hfill
  \begin{minipage}[t]{0.24\linewidth}
    \centering
    \includegraphics[width=\linewidth]{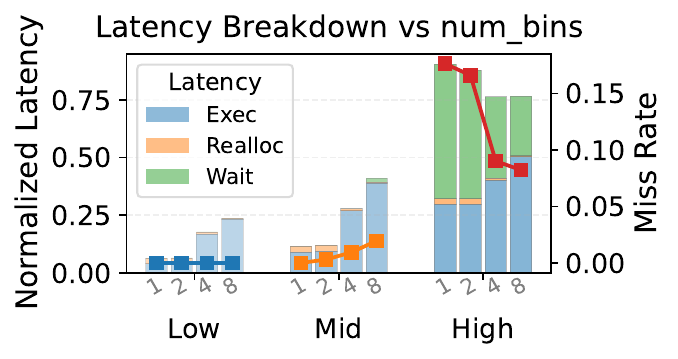}\\
    \small (c) Partitioning: latency breakdown
  \end{minipage}
  \hfill
  \begin{minipage}[t]{0.24\linewidth}
    \centering
    \includegraphics[width=\linewidth]{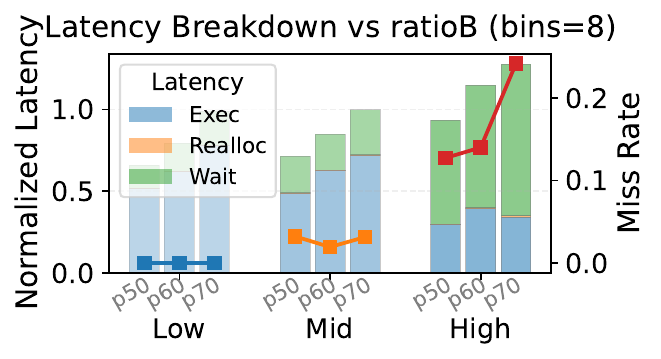}\\
    \small (d) Dynamic reservation + partitioning
  \end{minipage}
  \vspace{-0.5em}
  \caption{Ablation study. (a) Effect of dynamic reservation: \CyclicName(S) vs.\ \CyclicName{} (p80 projected as dashed lines). (b--c) Effect of spatial partitioning: rescheduling overhead and latency breakdown vs.\ partition count. (d) Effect of dynamic reservation under partitioning: latency breakdown vs.\ reservation percentile.}
  \label{fig::ablation}
  \vspace{-0.5em}
\end{figure*}

To verify the effectiveness of our proposed mechanisms, we conduct three ablation experiments that isolate the effects of reservation and spatial partitioning.
Figure~\ref{fig::ablation} presents the results across all experiments.

\subsubsection{Effect of Dynamic Reservation}
\label{sec::abla_reservation_serial}
\revision[blue][R2-Q7]{%
We first isolate the effect of \emph{dynamic reservation} by comparing \CyclicName{} (hard sub-deadlines) with its elastic variant \CyclicName(S).
The key difference is that \CyclicName{} enforces fixed deadlines, while \CyclicName(S) uses expected ready time (ERT) and deadline (DDL) as elastic constraints that allow E2E slack sharing across the \ETEs chain.
Both use identical spatial partitions and tile budgets.
Figure~\ref{fig::ablation}(a) shows the utilization--reliability tradeoff as we sweep the quantile $q$ from p50 to p70.%
}

Two observations emerge.
First, \CyclicName(S) at $q{=}$p60 achieves a miss rate of 6.0\%, lower than \CyclicName{} at p80 (8.1\%), demonstrating that E2E slack sharing improves reliability at the same resource budget.
Second, \CyclicName(S) reduces the idle ratio from ${\sim}33\%$ to ${\sim}20\%$---a 39\% reduction---confirming that the gains come from better slack redistribution.
\CyclicName(S) achieves significant utilization improvement and miss rate reduction at minimal rescheduling overhead ($<$0.4\%).

\subsubsection{Effect of Spatial Partitioning}
\label{sec::abla_isolation}
We next isolate the effect of \emph{spatial partitioning} by comparing \TpFirstName{} ($\PartitionNum{=}1$) with its partitioned variant ($\PartitionNum \in \{2, 4, 8\}$).
We also sweep hardware scale and load scale: (Light, Medium, and Heavy) with tile count (400, 400, 200) and load factor (0.5, 1.0, 1.0).
Figure~\ref{fig::ablation}(b) shows that as $\PartitionNum$ increases from 1 to 8, the rescheduling overhead (realloc.\ ratio) drops sharply (20.2\%$\to$2.0\% under Low load, 6.9\%$\to$0.4\% under High load) while the number of rescheduling events ($N_{Rch}$) remains largely unchanged.
This confirms that partitioning reduces per-switch cost by localizing reallocation within each partition.

Figure~\ref{fig::ablation}(c) reveals the accompanying tradeoff.
Under High load, increasing $\PartitionNum$ from 1 to 8 reduces miss rate from 23.5\% to 7.8\% because isolation prevents interference cascades.
Under Low and Mid loads, additional partitions slightly increase idle time due to reduced cross-partition sharing.

\subsubsection{Effect of Dynamic Reservation under Partitioning}
\label{sec::abla_reservation_parallel}
Finally, we examine how dynamic reservation behaves under partitioning by comparing pglb with reserv (both using 8 partitions).
Figure~\ref{fig::ablation}(d) reveals a striking contrast with the unpartitioned case.
Whereas dynamic reservation monotonically improves miss rate in \CyclicName(S), we observe a non-monotonic U-shaped trend under partitioning.
Under High load, miss rate rises from 15.5\% (p50) to 22.8\% (p70), exceeding the pglb baseline of 7.8\%.
Under Mid load, a sweet spot emerges at p60 (2.8\%).
This behavior stems from the interplay between slack redistribution and intra-partition contention.
When multiple tasks within a partition have tight ERT/DDL constraints, the scheduler triggers frequent preemption to honor them.
The net effect depends on load intensity: under heavy contention, preemption overhead outweighs slack redistribution benefits; under light contention, dynamic reservation has little impact.

This result highlights a key design insight: \emph{dynamic reservation and spatial partitioning must be jointly tuned}---neither mechanism alone suffices for optimal performance.

\paragraph{Guidelines for quantile selection.}
\revision[blue][R4-Q3]{%
The above ablation reveals that the optimal $q$ depends on load intensity and the degree of spatial partitioning.
We recommend the following two-step approach:
\begin{enumerate*}[label=(\roman*)]
  \item Start with a conservative $q$ (e.g., p99) as the upper bound, derived from the target E2E reliability specification;
  \item Relax $q$ progressively (e.g., to p95 or p90) while monitoring the actual E2E miss rate, exploiting the tail-composition headroom discussed in Section~\ref{sec::latency_model} and the slack-sharing mechanism (Section~\ref{sec::hybrid-grained_isolation}).
\end{enumerate*}
In practice, for workloads with moderate variation, setting $q$ to p95--p98 provides sufficient reliability margin while reducing tile demand by 10--15\% compared to a naive p99 baseline.%
}

\subsection{End-to-End Evaluation}%
\label{sec::exp_ETEs}

\subsubsection{Evaluation Metrics} 
% To evaluate the performance effectiveness of our proposed solution, we use the following metrics:
We compare our method against baseline methods in the following respects:
\begin{enumerate}[leftmargin=0pt, itemindent=2.0em, labelsep=0.5em, label=(\arabic*)]
  % \item \textbf{Latency violation rate}: the percentage of queries that exceed the latency bound. 
  %     This metric is used to measure the scheduling performance in job granularity. 
  %     Even though ADS does not enforce hard time constraints~\cite{Luo2019-UCB_safeos} for every job,
  %     A miss or delay in the execution of any module may not lead timeout of the final output, 
  %     the control signal will be dropped, and the screen will endurance 
  %     video image frame to distort or freeze and will result in passenger distrust.
  \item \option{Latency violation rate}: the percentage of \ETEs chain instances that exceed the latency bound. 
          This metric is measured at the \ETEs level because slack is shared along the task chain.
  \item \option{99th-percentile tail latency}: the E2E tail-latency bound at the 99th percentile (p99).
            This metric measures E2E performance when timed-out chain instances are not dropped. 
  \item \option{Scaling efficiency}: 
        We evaluate the scaling efficiency of our method by comparing (1) the resource capacity requirement to meet the latency bound for a given workload scale, and (2) the maximum workload scale that can be accommodated with no timeout under a specific hardware capacity. We scale all resource capacity linearly with the number of tiles; for simplicity, we use $N_{\Tile}$ to indicate the resource capacity.
  % \item Jitter per frame: the average per-frame execution time variation at the \ETEs level and the functions level~\cite{Alcon2020}.
\end{enumerate}

\begin{figure*}[htp]
  \centering
  \includegraphics[width=0.95\linewidth]{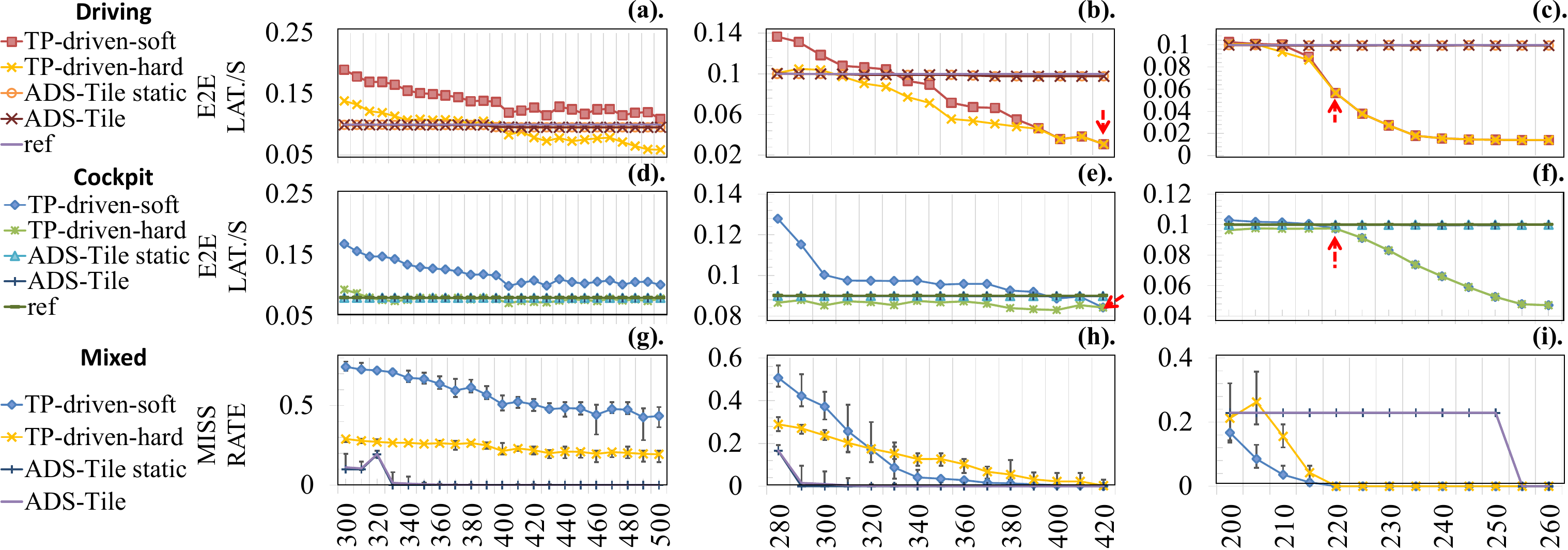}
  \caption{The 99th-percentile E2E tail latency and deadline miss rate, 
  with respect to the number of tiles, under different workloads (i.e., number of \aux chains and \ETEs deadline): light ($\times 1, 100$\,ms), 
  medium ($\times 6, 90$\,ms), and heavy ($\times 9, 80$\,ms), 
  and different drop policies (i.e., hard and soft). 
  }%
  \label{fig::exp_latency}
  \vspace{-0.5cm}
\end{figure*}
Figure~\ref{fig::exp_latency} shows the 99th-percentile E2E tail latency (a-f), 
and the latency violation rate (g-i). 
We evaluate three representative cases: light (a, d, g), medium (b, e, h), and heavy (c, f, i) workloads,
and compare the performance of our method with \revision[blue][R2-Q8]{\TpFirstName (represented by Planaria~\cite{Ghodrati2020-Planaria})}.
Driving and \aux tail latencies are plotted separately in (a-c) and (d-f), respectively.
% Considering the mechanism that the scheduler will drop the timeout tasks,
% we also test the case that the timeout tasks are not dropped (hard).
To handle timeout events (chains), we test two mechanisms:
drop the timed-out event (hard) or allow it to continue executing (soft).

\subsubsection{Latency violation rate}
As shown in Figure~\ref{fig::exp_latency} (g-i),
the violation rate decreases as the number of tiles increases for both task types
and across all three workloads.
Under workload variation and hardware contention, \PaperName keeps the 99th-percentile tail latency close to the no-variation baseline while requiring far fewer tiles (300 vs.\ 400) to meet the latency bound, as shown in Figure~\ref{fig::exp_latency} (h-i).
Across repeated statistical trials, its deadline-miss rate also has lower variance than \TpFirstName, indicating more stable latency behavior without making a statistical-significance claim.

\subsubsection{Tail latency}
% 410 and 310 cores are required to meet the latency bound for all tasks in \TpFirstName and our method, respectively. 
% And a basic trend that the tail latency decreases as the number of tensor cores increases, can be observed in two kinds of tasks.
% However, our method shows a relatively flat curve, as the number of tensor cores changes. 
% \TpFirstName requires 410 tensor cores to meet the latency bound for all tasks (sink nodes at the end of each chain)
% and shows an obvious trend that the tail latency decreases as the number of tensor cores increases: 
The tail latency drops quickly in \TpFirstName as the number of tiles increases, from 130\,ms to 30\,ms for driving tasks and from 130\,ms to 80\,ms for \aux tasks. By contrast, our method presents a flatter curve, 
near the deadline bound in all cases.
This characteristic meets ADS requirements:
predictable and stable latency within the latency bound 
is more important than a lower tail latency. 
This difference stems from the design logic of \PaperName:
 avoid unnecessary acceleration and preserve resources for the most critical tasks.

 \subsubsection{Comparison with dropping techniques} 
%  \cite{Seah24-DREAM}
Some recent systems~\cite{Seah24-DREAM, Gog2022-Eurosys-D3} adopt dropping mechanisms that skip some tasks
to improve the tail latency of the remaining tasks when the scheduler cannot handle workload peaks with the available resources. 
We also compare our method with these dropping techniques. 
% Genetically, a bad scheduler will quickly exhaust the available resources under burst loads, causing long queue delays and timeouts. 
% When the scheduler fails to handle the peaks, the coming loads may overflow the available resources, two policies deviate from each other: 
Figure~\ref{fig::exp_latency} (a-f) shows two dropping policies on \TpFirstName:
\begin{enumerate*}[label=(\roman*)]
  \item Timing correctness first (Hard): timeout tasks are dropped once they exceed their sub-deadline, and downstream tasks reuse the stale data received in the last period. 
  \item Algorithm correctness first (Soft): no timeout tasks are dropped, and downstream tasks keep waiting for the correct data version.
\end{enumerate*}
As can be seen in Figure~\ref{fig::exp_latency} (a-f), dropping timed-out tasks greatly reduces the overall latency of {\TpFirstName}.
As the hardware capacity increases, the benefits of dropping decrease, and 
the curves with hard and soft dropping policies get closer and finally intersect at the positions marked by red arrows. 
By contrast, \PaperName shows flat tail latency without a dropping policy and remains close to the no-variation baseline for workload and hardware contention. However, dropping timed-out tasks can cause control-frame drops or force downstream stages to act on stale data, reducing the reliability of downstream decisions in safety-critical paths.

% \takeaway{}

 \subsubsection{Inter-task interference}
Figure~\ref{fig::exp_latency} (c, f) and (a, d) show the results under light and heavy workloads,
and our method outperforms \TpFirstName in the heavy case but performs worse in the light case.
This results from resource partitioning in our method,
which forbids cross-partition allocation to control per-rescheduling cost under high load;
however, partitioning also sacrifices some sharing opportunities, requiring larger-scale hardware to provide the redundancy needed for isolation.
When the workload is light, the rescheduling overhead is small;
unconstrained resource sharing improves utilization. However, 
as workload scales, rescheduling overhead quickly becomes the bottleneck and overwhelms the benefit of resource sharing. 
This case shows that the best partitioning and mapping schemes depend on the workload and hardware capacity.

\subsubsection{Scaling performance}
\begin{figure}[ht]
  \centering
  \includegraphics[width=0.9\linewidth]{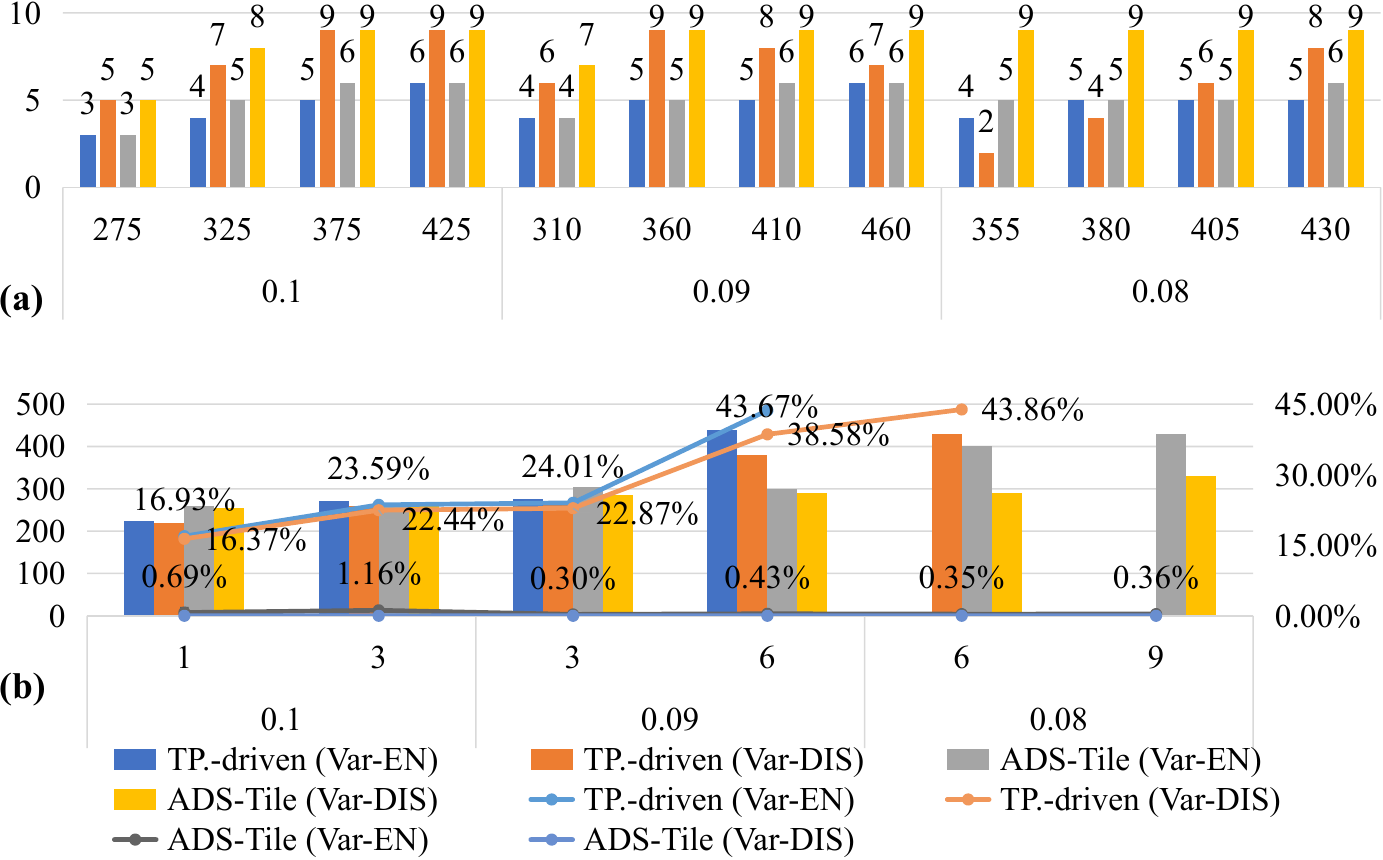}
  \caption{(a) The maximum number of \aux chains that can be accommodated with no timeout,
  (b) Required minimum number of tiles to meet the latency bound.}%
  \label{fig::exp_throughput}
  % \vspace{-0.5cm}
\end{figure}
 Figure~\ref{fig::exp_throughput} (a) and (b) show the scaling performance
 concerning hardware scale and workload scale, respectively.
 Figure~\ref{fig::exp_throughput} (a) shows the maximum number of \aux chains
 for each hardware configuration with runtime variability enabled (Var-EN) or disabled (Var-DIS), 
 under the constraint that all tasks meet the latency bound.
 First, the number of supported \aux chains increases as the number of tiles scales up.
 However, \TpFirstName cannot accommodate more than $\times 6$ \aux chains, 
 and \PaperName increasingly outperforms \TpFirstName. 
 Overall,
 \PaperName accommodates $1.3$--$2.3\times$ more \aux chains than \TpFirstName.
 Second, it can be observed that the difference between the 
  two cases (EN and DIS) is significant in 
  the case of 355 tiles and 80\,ms latency bound, 
  and this gap diminishes as resources increase. 
  For a given workflow graph, when the total number of tiles is limited,
  fine-grained partitioning cannot guarantee that all tasks fit within their assigned partitions;
  consequently, even critical tasks must endure significant intra-partition resource contention and rescheduling.
  As the number of tiles increases, this problem is gradually alleviated.
 Figure~\ref{fig::exp_throughput} (b) compares the minimum tile count
  required to meet each deadline.
  \PaperName consistently needs fewer tiles than \TpFirstName, and the gap widens as workload scales:
  under the heaviest configuration, \TpFirstName cannot meet the deadline at any tested capacity.
  We quantify the resulting hardware cost reduction below.
%   璐熻浇浣庯紝var浣?鈫?any, % 璐熻浇浣庯紝var楂?鈫?glb, % 璐熻浇楂橈紝var浣?鈫?cyc, % 璐熻浇楂橈紝var楂?鈫?Ours
    These results indicate that the best scheduling scheme is not always the one with the lowest latency bound. 
    Depending on the workload characteristics and the available resources, the best scheduling scheme may vary: 
  {\CyclicName} is well-suited for heavy and low-variance workloads,
  whereas \TpFirstName is designed for lightweight and high-variance workloads.
  Lastly, \PaperName performs best in heavy, high-variance workloads.

\revision[blue][R3-Q4]{%
\textbf{Hardware cost reduction.}
Figure~\ref{fig::exp_throughput}(b) directly answers the question ``how many tiles are needed to meet a given E2E deadline?''
Under medium load ($\times 6$ \aux, 90\,ms deadline), \TpFirstName requires 440 tiles whereas \PaperName needs only 300 tiles---a 31.8\% tile reduction.
Under heavy load ($\times 6$ \aux, 80\,ms; or $\times 9$ \aux, 80\,ms), \TpFirstName cannot meet the deadline even with 500+ tiles, while \PaperName succeeds with 400 and 430 tiles respectively.
Moreover, the scheduling-induced overhead (cum.\ wasted processing power) drops from 17--44\% under \TpFirstName to below 1.2\% under \PaperName across all configurations, meaning nearly all allocated tile capacity is spent on useful computation.
In lighter configurations ($\times 1$ \aux, 100\,ms), \PaperName requires marginally more tiles (260 vs.\ 225) due to partition fragmentation, but its cumulative overhead remains 24$\times$ lower (0.69\% vs.\ 16.93\%), yielding more predictable latency.
Overall, \PaperName reduces tile demand by up to 32\% under deadline-critical workloads, translating directly to proportional silicon area and power savings.}

\subsection{Scheduling Overhead}
\label{sec::sched_overhead}

\revision[blue][R3-Q5]{%
Algorithm~\ref{alg::multi_chain_slack_distribution} (multi-chain slack distribution) executes offline during compile time and thus is not on the critical path.
Algorithm~\ref{alg:reschedule} (dynamic colocation and allocation) is triggered at runtime when reallocation is needed. The overall reallocation overhead consists of the scheduling decision latency and the data resharding latency.
}

\begin{table}[ht]
  \centering
  \caption{Runtime overhead of Algorithm~\ref{alg:reschedule}.}
  \label{tab::sched_overhead}
  \footnotesize
  \setlength{\tabcolsep}{4pt}
  \begin{tabular}{lcccc}
    \toprule
    \multirow{2}{*}{Configuration} & \multicolumn{4}{c}{Ratio (\%)} \\
    \cmidrule(lr){2-5}
    & Mean & P50 & P99 & Max \\
    \midrule
    1 partition (glb)   & 7.7 & 5.5 & 32.7 & 45.1 \\
    4 partitions (pglb) & 4.6 & 2.8 & 21.0 & 69.2 \\
    \bottomrule
  \end{tabular}
\end{table}

\revision[blue][R3-Q5]{%
Table~\ref{tab::sched_overhead} reports the ratio of a single scheduling-decision latency to a triggered data resharding latency.
The mean scheduling-decision overhead is 7.7\% of data resharding latency for the single-partition case and 4.6\% for the multi-partition case. This confirms that scheduling-decision latency is not the dominant component of reallocation overhead.%
}

% \clearpage
\section{Related Work} \label{sec::related_work}

% 文章定位：Autonomous driving system，DNN-based Workflow acceleration, 
% Tile-based accelerator, low-level scheduler, \ETEs latency gurantee

% 本文研究了一个交叉topics的工作，即：
% 在下一代支持资源隔离和多任务并行的ADS计算平台上，如何高效的调度自动驾驶端到端的DNN工作流，
% runtime dynamics no-awaring 
% enforcement by accuracy drop
% spatial adaptation faces trivial switching cost 

% isolation mechanism
% temporal isolation
% budget isolation in time dimension

% that explores how to efficiently deploy DNN-based ADS workflows on next-generation ADS computing platforms that support resource isolation and multi-task colocation. 
This work targets an underexplored gap between ADS software scheduling and modern tile-based chips: real-time scheduling must provide \ETEs workflow latency guarantees, while tile-based DNN accelerators must support dense multi-task colocation under physical resource constraints.
In both contexts, we distinguish prior work by its underlying hardware model and the way it handles workload variation.

\begin{table}
    \centering
    \caption{Comparison of Different Schedulers. \vspace{-1mm}}
    \label{tab::related_works}
    \resizebox{\linewidth}{!}{%
    \begin{tabular}{l|c|c|c|c|c|c} 
        \toprule
        \multirow{2}{*}{Scheduler} & \multicolumn{2}{c|}{Workload}                         & Metric                    & \multicolumn{3}{c}{Feature}                                                        \\ 
        \cline{2-7}
                                                       & \ETEB                     & Variation                 & Deadline                  & Dynamic                   & Spatial                   & Isolation                  \\ 
        \hline
        ROS~\cite{Casini2019, valigi2021lessons}                                                                                                                   & \checkmark & \Xcross    & \Xcross    & \Xcross    & \Xcross    & \Xcross     \\ 
        \hline %Bohren2020-ROS2,Arafat2022-RT-analysis-ROS2,Casini2019,Choi21-PiCAS,CyberRT, ROS1   
        \begin{tabular}[c]{@{}l@{}}Reservation\\~\cite{baker1989cyclic, McLean2020-tttech_tablescheduling}\end{tabular}                                                               & \checkmark & \checkmark & \checkmark & \Xcross    & \Xcross    & \checkmark  \\ 
        \hline %Igarashi2019-multirate, Xu2022-AOI, Hsiao2022-Zhuyi
        \begin{tabular}[c]{@{}l@{}}lat.-Acc. trade-off\\~\cite{Gog2022-Eurosys-D3,Seah24-DREAM}\end{tabular}                                                                            & \checkmark & \checkmark & \checkmark & \checkmark & \Xcross    & \Xcross     \\ 
        \hline %Yao2020imprecise, Zheng2023TileFlow,
        \begin{tabular}[c]{@{}l@{}}Tile-based offline\\~\cite{Timeloop, huang2021cosa, Kao2022-MAGMA}\end{tabular}                                     & \Xcross    & \Xcross    & \Xcross    & \Xcross    & \checkmark & \Xcross     \\ 
        \hline % Atomic, Zheng23-COMB,Inter-layerSched,Hong2023-DOSA, 
        \begin{tabular}[c]{@{}l@{}}Tile-based serving\\~\cite{Ghodrati2020-Planaria,Liu2022-VELTAIR,Kim2023MoCA}\end{tabular} & \Xcross    & \Xcross    & \Xcross    & \checkmark & \checkmark & \Xcross     \\ 
        \hline
        \PaperName                                                                                                                                                                                                          & \checkmark & \checkmark & \checkmark & \checkmark & \checkmark & \checkmark  \\
        \bottomrule
    \end{tabular}  
      }
    % \vspace{-1.5em}
\end{table}

% ros + linux kernel
\paragraph*{Real-time schedulers for \ETEs workflow latency guarantees}
% ~\cite{Casini2019} 
Real-time schedulers primarily focus on time-slice management within homo-/hetero-multiprocessor systems. ROS, a popular middleware for robotics applications, is generally not suitable for real-time usage~\cite{Casini2019} due to its limited priority support and non-determinism in scheduling. Crucially, ROS functions merely as a middleware, lacking direct hardware resource management capabilities and relying on the underlying operating system. 
To handle workload variations, approaches such as SCHED\_DEADLINE~\cite{Lelli2016} and cyclic schedulers~\cite{McLean2020-tttech_tablescheduling} adopt reservation (temporal isolation)~\cite{Casini2019,Lelli2016}. In this paradigm, each processor is pinned to a group of periodic tasks, and within each period, tasks are exclusively reserved with a fixed budget of processing time. This mechanism prevents excessive consumption of processing time, thereby isolating cascading timeouts that occur when tasks are time-multiplexed on the same core. However, such fixed per-task reservation is often overly conservative, leading to significant waste of processing time. 

% and others~\cite{Hsiao2022-Zhuyi}
To mitigate this waste, systems such as D3~\cite{Gog2022-Eurosys-D3} and DREAM~\cite{Seah24-DREAM} reduce these budgets and trade off latency guarantees with algorithmic accuracy. This is typically achieved through:
\begin{enumerate*}
\item Implementation techniques such as DNN early exit, which enable sending an inaccurate result with lower latency.
\item Task dropping when overload occurs, to guarantee that the remaining tasks are satisfied.
\end{enumerate*}
However, accuracy loss is not acceptable in all traffic conditions.
\revision[blue][R1-Q1]{Dynamic reservation servers like CBS~\cite{Lelli2016} and GRUB~\cite{lipari2000grub} reclaim unused temporal budgets at runtime to improve utilization while preserving probabilistic latency guarantees.
Jigsaw~\cite{sun2024jigsaw} implements a similar idea for ADS on a multi-accelerator architecture, sharing the unused time budget of critical tasks with non-critical tasks.
However, these methods assume fixed processing units and cannot leverage spatial tunability.
\PaperName extends this probabilistic guarantee to the spatial dimension ($c_v$), making the quantile $q$ a joint space-time knob for tile-based accelerators.}

\paragraph*{\HWName accelerators for multi-DNN colocation} 
No real-time work explores isolation granularity on spatial resources, which is crucial for state-of-the-art L4+ ADS.
\HWName accelerators bridge this gap, with compute, memory, and control resources spatially partitioned to enable efficient isolation and colocation of multiple DNN tasks on different tiles.
Some works have studied the static mapping problem,
which seeks the optimal mapping for
dependent-task~\cite{Kao2022-MAGMA}, layer (i.e., operator), 
and nested-loop~\cite{Timeloop, huang2021cosa} levels to maximize data reuse and minimize communication cost. However, these works do not consider dynamic workload changes, leading to resource waste or 
missed deadlines. 
More recently, several works have considered serving DNNs with QoS constraints~\cite{Ghodrati2020-Planaria} on \HWName accelerators. They adaptively reallocate spatial resources among co-located DNNs to improve utilization during non-busy periods and reduce response latency during busy periods. However, these methods overlook interference among co-located DNNs, which primarily includes scheduling overhead and contention for shared system resources.

\begin{enumerate*}[label=]
\item V10~\cite{Xue2023V10} focuses on imbalanced use of vector and matrix units within the tiles, rather than general system resource contention.
\item VELTAIR~\cite{Liu2022-VELTAIR} considers both and employs coarse-grained layer-block scheduling to reduce the frequency of resource reallocation while exploring compilation schemes that are less sensitive to shared resource contention.
\item MoCA~\cite{Kim2023MoCA} adopts a hardware-software co-design to support dynamic, contention-aware bandwidth repartitioning.
\item AuRORA~\cite{kim2023aurora} introduces a virtualization layer to manage the spatial mapping from logical to physical resources and provides a scheme that minimizes switch overhead for every single reallocation operation.
\end{enumerate*}
However, these methods are not suitable for ADS due to the frequent reallocation caused by the high task arrival rate in ADS, which seriously impacts \ETEs latency.

% However, none of methods is fully suitable for ADS. First, they primarily focus on throughput under work-conserving scheduling, greedily maximizing resource utilization at all times; this leads to frequent spatial reallocation, whose cumulative scheduling-induced overhead is unacceptable to \ETEs systems like ADS.
% Secondly, they lack a unified isolation-sharing space abstraction in space and time. Finally, they can not proactively avoid interference among co-located DNNs from a perspective of \ETEs workflow. 
% Our \PaperName framework is specifically designed to fill these gaps. 

\section{Conclusion}%
\label{sec::conclusion}
% Next-generation autonomous driving systems (ADS) adopt tile-based computational platform that stress the colocation of multiple ADS on single chips for lower cost, as well as using programmable isolation capacity to isolate interference among the co-running tasks. 
% The interdisciplinary real-time system (ADS) and emerging tile-based ADS platform that is overlooked by the existing scheduling methods. 
This paper presents \PaperName, an isolation-aware scheduling framework for tile-based ADS platforms.
The scheduling problem is spatio-temporal: changing DoP can free tiles for other tasks, but frequent stop-migrate-restart reallocation uses part of the E2E latency budget.
\PaperName combines configurable isolation and elastic reservation to create a spatio-temporal isolation-sharing space.
Within this space, the runtime scheduler shares resources across the DAG while limiting reallocation in both space and time.
Across our benchmarks, \PaperName reduces tile demand and wasted processing capacity from reallocation, and improves deadline satisfaction compared with baseline schedulers.

\bibliographystyle{IEEEtran}
\bibliography{IEEEabrv,./ref}
% \bibliography{./ref}
\vspace{-33pt}
\begin{IEEEbiographynophoto}{Chenguang Zhang}
received the B.S. degree in electronic engineering from Tianjin University, Tianjin, China, in 2016, and the master's degree in computer science from ShanghaiTech University, Shanghai, China, in 2021. He is now pursuing the Ph.D. degree in the School of Computer Science at Peking University. His current research interests include spatial architectures, reliability-aware design methodologies, algorithm-hardware co-design, and design space exploration.
\end{IEEEbiographynophoto}

\vspace{-33pt}
\begin{IEEEbiographynophoto}{Yuanpeng Zhang}
    received the B.S. degree in computer science from Peking University, Beijing, China, in 2022. He is now pursuing the Ph.D. degree in the School of Integrated Circuits at Peking University. His research interests include NN inference, high-performance PIMs, and software-hardware co-design for accelerators.
\end{IEEEbiographynophoto}

\vspace{-33pt}
\begin{IEEEbiographynophoto}{Chenhao Xue}
received the B.S. degree in computer science from Peking University, Beijing, China, in 2023. He is now a Ph.D. candidate in the School of Integrated Circuits, Peking University. His current research interests include system-technology co-optimization (STCO), design automation, and domain-specific accelerators.
\end{IEEEbiographynophoto}

\vspace{-33pt}
\begin{IEEEbiographynophoto}{Yihan Yin}
    received the B.S. degree in physics from Peking
    University, Beijing, China, in 2025. He is now a Ph.D. candidate in the School
    of Integrated Circuits, Peking University.
\end{IEEEbiographynophoto}

% \vspace{-33pt}
% \begin{IEEEbiographynophoto}{Bo Yu}
% xxx

% \vspace{-33pt}
% \begin{IEEEbiographynophoto}{Shaoshan Liu}
% xxx

\vspace{-33pt}
\begin{IEEEbiographynophoto}{Chen Zhang}
(Member, IEEE) is a Tenure-Track Assistant Professor with Shanghai Jiao Tong University. Previously, he received the Ph.D. degree in EECS from Peking University in 2017 and then served as a senior researcher at Microsoft Research and a GPGPU architect at Alibaba. His research interests include computer architectures and heterogeneous computing for cloud \& edge AI systems. He received the FPGA 2015 Best Paper Nomination, the TCAD 2019 Donald O. Pederson Best Paper Award, and other honors.
\end{IEEEbiographynophoto}

\vspace{-33pt}
\begin{IEEEbiographynophoto}{Guangyu Sun}
is currently an Associate Professor in the School of Integrated Circuits at Peking University. He received his B.S. and M.S. degrees from Tsinghua University, Beijing, in 2003 and 2006, respectively, and his Ph.D. degree from the Pennsylvania State University in 2011. His research interests include design and automation for computer architecture, cross-layer co-optimization, and emerging memory technologies. He has published more than 150 journal articles and refereed conference papers in venues such as ISCA, MICRO, HPCA, DAC, and IEEE TCAD. His work has been recognized with the DAC Under-40 Innovators Award, the CCF-IEEE CS Young Computer Scientists Award, the Microsoft Research Asia Collaborative Research Award, the CCF-Intel Young Faculty Researcher Program, and four best paper awards. 
% He was the general co-chair of NVMSA2021 and the TPC co-chair of NVMSA2020, RTCSA2019, APPT2017, and NAS2012. 
He has served as a program committee member and a track chair for over 20 conferences in these areas, including DAC, ICCAD, MICRO, HPCA, etc. He is an associate editor of ACM JETC.
\end{IEEEbiographynophoto}

%%%%%%%%%%%%%%%%%%%%%%%%%%%%%%%%%%%%
% \appendix
% \clearpage
% \input{tex/4_techniques/sched_example.tex}

\end{document}